\documentclass[10pt]{iopart}
\usepackage{graphicx,bm,amssymb}
\eqnobysec

\usepackage[colorlinks,linkcolor=blue,urlcolor=black,citecolor=blue]{hyperref}

\newcommand{\beq}{\begin{equation}}
\newcommand{\eeq}{\end{equation}}
\newcommand{\beqa}{\begin{eqnarray}}
\newcommand{\eeqa}{\end{eqnarray}}

\newcommand{\abs}[1]{{\left\vert#1\right\vert}}
\renewcommand{\bar}[1]{{\overline{#1}}}
\newcommand{\dd}{{\rm d}}
\newcommand{\del}{{\bm\delta}}
\newcommand{\erf}{{\mathop{\rm erf}}}
\newcommand{\half}{{\scriptstyle1\over\scriptstyle2}}
\newcommand{\mean}[1]{{\langle#1\rangle}}
\newcommand{\prob}{\mathbb{P}}
\newcommand{\rad}{{({\rm rad})}}
\newcommand{\stir}[2]{\left[#1\atop#2\right]}
\renewcommand{\th}{{\theta}}
\newcommand{\var}{{\mathop{\rm var}\,}}
\newcommand{\x}{{\bm x}}
\newcommand{\y}{{\bm y}}
\newcommand{\area}{{\cal A}}
\newcommand{\A}{{\bm A}}
\renewcommand{\b}{{\bm b}}
\newcommand{\C}{{\bm C}}
\newcommand{\R}{{\bm R}}
\newcommand{\0}{{\bm 0}}

\begin{document}

\title{On sequences of convex records in the plane}

\author{Claude Godr\`eche and Jean-Marc Luck}

\address{Universit\'e Paris-Saclay, CNRS, CEA, Institut de Physique Th\'eorique,
91191~Gif-sur-Yvette, France}

\begin{abstract}
Convex records have an appealing purely geometric definition.
In a sequence of $d$-dimensional data points,
the $n$-th point is a convex record if it lies outside the convex hull of all preceding points.
We specifically focus on the bivariate (i.e., two-dimensional) setting.
For iid (independent and identically distributed) points,
we establish an identity relating
the mean number $\mean{R_n}$ of convex records up to time $n$
to the mean number $\mean{N_n}$ of vertices in the convex hull of the first $n$ points.
By combining this identity with extensive numerical simulations,
we provide a comprehensive overview of the statistics of convex records
for various examples of iid data points in the plane:
uniform points in the square and in the disk,
Gaussian points and points with an isotropic power-law distribution.
In all these cases, the mean values and variances of $N_n$ and $R_n$
grow proportionally to each other,
resulting in finite limit Fano factors $F_N$ and $F_R$.
We also consider planar random walks, i.e., sequences of points with iid increments.
For both the Pearson walk in the continuum and the P\'olya walk on a lattice,
we characterise the growth of the mean number $\mean{R_n}$ of convex records
and demonstrate that the ratio $R_n/\mean{R_n}$ keeps fluctuating
with a universal limit distribution.
\end{abstract}

\eads{\mailto{claude.godreche@ipht.fr},\mailto{jean-marc.luck@ipht.fr}}

\maketitle

\section{Introduction}
\label{intro}

The statistics of rare and extreme events are of great importance across various scientific disciplines.
In particular, the study of the statistics of records in discrete time series
has widespread relevance in fields such as climate studies, finance and economics, hydrology,
sports, and complex physical systems\footnote{An extensive list of references
on applications of records to the fields mentioned above can be found in~\cite{revue}.}.
Most of these studies deal with \emph{univariate} records.
The broad scope of the present work is to revisit the subject of \emph{multivariate} records.

Consider an infinite sequence of data points
$\x_1,\x_2,\dots,\x_n,\dots$ in $d$-dimensional space.
The label~$n$ will be referred to as a discrete time.
Loosely speaking,
$\x_n$ is a record whenever it is, in some sense, larger than all previous data points.
We then say that there is a record-breaking event, or a record for short, at time $n$.

In the one-dimensional or univariate setting, the data $x_n$ consists of real numbers.
There is a natural ordering of points on the line,
and therefore a natural definition of (upper) records.
There is a record at time $n$ if
\beq
x_n>\max(x_1,\dots,x_{n-1}).
\eeq
This canonical definition of univariate records
is invariant under the action of a large group of reparametrisation transformations.
Records are indeed left unchanged if the data $x_n$ are transformed into $y_n=y(x_n)$,
where $y(x)$ is any continuous increasing function.
The theory of univariate records,
initiated by Chandler~\cite{chandler} and R\'enyi~\cite{renyi1,renyi2},
has become a mature subject~\cite{glick,arn,nevzorov,nevzorov2,bunge}.
Most classic results concern the case where the $x_n$ are iid
(independent and identically distributed) continuous random variables.
The key property of records for iid variables is that there is a record at time~$n$
with probability
\beq
Q_n=\frac{1}{n},
\label{pusuel}
\eeq
independently of other occurrences of records.
This result holds irrespective of the underlying distribution of the random variables $x_n$,
provided the latter is continuous.
The resulting distribution of the number of records $R_n$ up to time $n$ has been long known
(see~\ref{app}).
Another well-studied case is when the data points are the successive positions
of a one-dimensional random walker.
Now, the data points are no longer iid, but their increments are.
The basic knowledge on records for random walks can be found in~\cite{feller2}.
This problem has later been revisited in the physics literature
(for a review, see~\cite{revue} and references therein).

In the multivariate setting, the data points $\x_n$ are $d$-dimensional vectors.
At variance with the one-dimensional situation,
there is no natural total ordering in $d$-dimensional space.
This observation was made long ago~\cite{K66,B76}
and led to a variety of definitions of multivariate records~\cite{GR89,GR95}
(see also~\cite{arn,gnedin,HT10,DZ18,BSN20,TF21}).
One of these definitions stands out for
its minimalistic beauty---the notion of convex records, defined as follows.
There is a record at time $n$ if $\x_n$ does not belong to the
convex hull $\C(\x_1,\dots,\x_{n-1})$ of all previous data points,
i.e., the smallest closed convex set containing these points.
Convex records have attracted very little attention so far.
The phrase `convex records' with this meaning seems
to appear only once in the literature~\cite{K95}.
Introductions to the convex geometry of random sets can be found in~\cite{KM,SW}.
For a review on convex hulls in the physics literature, see~\cite{MCR}.

The purpose of the present work is to investigate the statistics of these convex records.
Henceforth we focus our attention on the planar case,
where data points~$\x_n$ are two-dimensional vectors representing points in the plane,
and the convex hull of the first $n$ points is a convex polygon.
We consider two different settings,
namely iid data points in section~\ref{iid}
and random walks in section~\ref{rw}.
The results to be described below make no claim to mathematical rigour.
They are based on a combination of analytical reasoning,
numerical simulations and heuristic scaling analysis.

One of the key geometric properties of the convex hull of a set of points
is that it is invariant under the action of the affine group.
In the bivariate setting of the present work,
if the data points $\x_n$ are changed into $\y_n=\A\x_n+\b$,
where~$\A$ is a constant invertible matrix (i.e., $\det\A\ne0$) and $\b$ a constant vector,
their convex hull is changed by the same affine transformation.
This symmetry is less constraining than the reparametrisation invariance of the univariate case.
It nevertheless gives rise to interesting consequences.
For instance, the statistics of convex records for uniform iid data points
in the unit square (resp.~in the unit disk)
is identical to that of uniform points in any parallelogram (resp.~in any ellipse).

\begin{figure}[!ht]
\begin{center}
\includegraphics[angle=0,width=0.56\linewidth,clip=true]{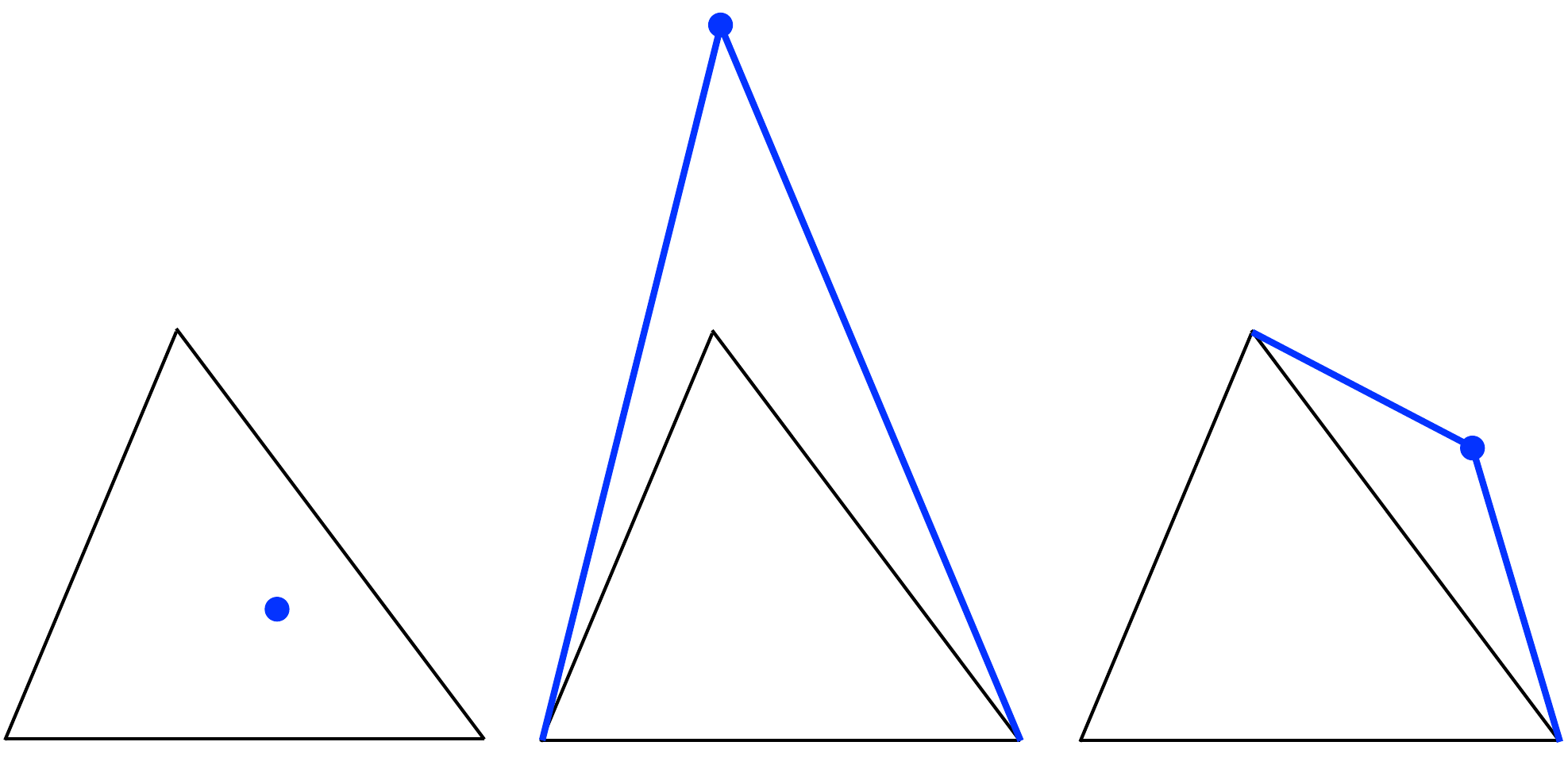}
\caption{\small
The three possible cases of convex records for $n=4$ (see~(\ref{nr4})).
The fourth data point and the attached edges of the convex hull, if any, are shown in colour.
Left to right: $(N_4=3,\,R_4=3)$, $(N_4=3,\,R_4=4)$, $(N_4=4,\,R_4=4)$.}
\label{34}
\end{center}
\end{figure}

\begin{figure}[!ht]
\begin{center}
\includegraphics[angle=0,width=0.4\linewidth,clip=true]{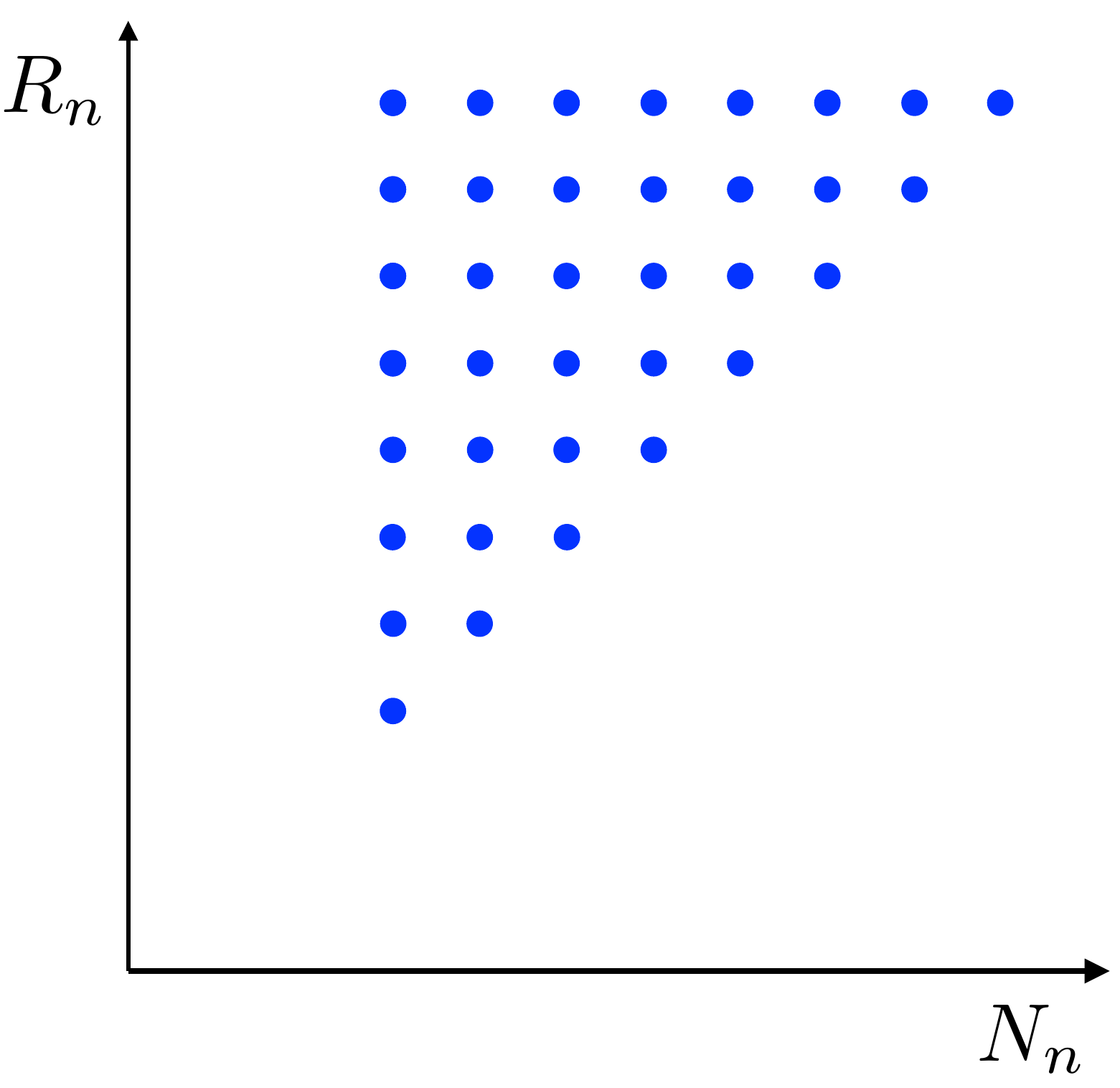}
\caption{\small
The 36 possible values taken by the couple $(N_n,R_n)$ for $n=10$.}
\label{NR}
\end{center}
\end{figure}

Let us introduce a few notations.
For a given sequence of two-dimensional data points $\x_n$,
we denote by $N_n$ the number of vertices of the convex hull of the first $n$ points,
and by $R_n$ the number of convex records up to time~$n$.
In the generic situation where the first three points are not aligned,
each of them is a convex record.
We have therefore
\beq
N_1=R_1=1,\quad
N_2=R_2=2,\quad
N_3=R_3=3.
\label{nr123}
\eeq
The construction begins to be non-trivial for $n=4$,
resulting in three possibilities (see figure~\ref{34}):
\beq
(N_4=3,\,R_4=3),\quad
(N_4=3,\,R_4=4),\quad
(N_4=4,\,R_4=4).
\label{nr4}
\eeq
More generally, $N_n$ and $R_n$ may take any values in the range (see figure~\ref{NR})
\beq
3\le N_n\le R_n\le n.
\label{nrtri}
\eeq
There are therefore $(n-1)(n-2)/2$ possible couples $(N_n,R_n)$.

The setup of this paper is as follows.
Section~\ref{iid} is devoted to sequences of iid data points.
We consider four characteristic examples,
namely uniform points in the square (section~\ref{square}),
uniform points in the disk (section~\ref{disk}),
Gaussian points (section~\ref{gauss}),
and points with an isotropic power-law distribution (section~\ref{levy}).
Section~\ref{sum} presents a summary on the mean values and variances of $N_n$ and $R_n$,
whereas the extremal probabilities that $N_n$ and $R_n$ take their smallest or largest values
are considered in section~\ref{ext}.
In section~\ref{rw} we investigate convex records of planar random walks.
The Pearson walk in the continuum and the P\'olya walk on three lattices are dealt with in parallel.
Section~\ref{disc} contains a brief discussion.
We recall the classical theory of univariate records in an appendix.

\section{Sequences of iid data points}
\label{iid}

This section is devoted to the situation where the data points $\x_n$
are iid and drawn from some arbitrary distribution in the plane,
assumed to be continuous, to prevent any ties.

\subsection{General results}

Let us denote by $I_n$ the characteristic function of the event
that there is a convex record at time $n$,
and by $Q_n=\mean{I_n}$ the corresponding record-breaking probability.
The random number $R_n$ of records up to time $n$ reads
\beq
R_n=\sum_{m=1}^n I_m.
\eeq
We have therefore
\beq
\mean{R_n}=\sum_{m=1}^n Q_m,
\label{rpro}
\eeq
and so
\beq
Q_n=\mean{R_n}-\mean{R_{n-1}}.
\label{pror}
\eeq

Furthermore, in the iid setting, data points are exchangeable.
Hence $Q_n$ is the probability that any point $\x_m$ chosen among the first $n$ ones
does not belong to the convex hull of all the other ones.
The product $n Q_n$ is therefore the mean number of points
among the first $n$ ones which are not inside the convex hull of all the other ones.
These points are precisely the vertices of $\C(\x_1,\dots,\x_n)$.
This translates~to
\beq
n Q_n=\mean{N_n}.
\label{pron}
\eeq
The univariate result~(\ref{pusuel}) is recovered by setting $\mean{N_n}=1$ in the above formula.
Eliminating the record-breaking probability $Q_n$ between~(\ref{pror}) and~(\ref{pron}),
we obtain the following identity:
\beq
\mean{N_n}=n\bigl(\mean{R_n}-\mean{R_{n-1}}\bigr).
\label{nriden}
\eeq

It is worth emphasising that the formulas~(\ref{rpro}) and~(\ref{pror}) hold in full generality,
whereas~(\ref{pron}) and~(\ref{nriden}) are specific to the case of iid data points.
Moreover,~(\ref{nriden}) only relates the mean value of $R_n$ to that of $N_n$.
The full statistics of $R_n$ is by no means simply related to that of $N_n$.

The first instance where the identity~(\ref{nriden}) is non-trivial is $n=4$.
This situation corresponds to Sylvester's famous four point problem~\cite{sylvester}
(see~\cite{P89} for a historical account).
Sylvester was interested in the probability $P$ that a random planar quadrilateral is reentrant, i.e., non-convex.
The exchangeability of the data points implies that the Sylvester probability reads
\beq
P=4q,
\eeq
where $q$ is the probability that a random point $\x$
is inside the triangle $(\x_1,\x_2,\x_3)$.
This reads formally
\beq
q=\mean{A(\x_1,\x_2,\x_3)},
\label{qdef}
\eeq
where $A(\x_1,\x_2,\x_3)$ is the `probability content' of the triangle $(\x_1,\x_2,\x_3)$,
i.e., the probability for a data point $\x$ to be inside that triangle,
and the average is taken over the three iid points $\x_1$, $\x_2$ and $\x_3$.
The probability $q$ depends on the underlying distribution of data points.
It has been known since the 19th century
for points uniformly distributed in various domains:
\beqa
&q({\rm triangle})=\frac{1}{12}=0.083333\dots,\quad
q({\rm square})=\frac{11}{144}=0.076388\dots,
\nonumber\\
&q({\rm disk})=\frac{35}{48\pi^2}=0.073880\dots
\eeqa
The triangle and the disk respectively yield upper and lower bounds of $q$
for points uniformly distributed in a domain.
Larger values of $q$ may however be reached.
We have indeed
\beq
q=1-\frac{3}{2\pi}\arccos\left(-\frac{1}{3}\right)=0.087739\dots
\eeq
for Gaussian points (see~(\ref{nraves4}),~(\ref{gn4})).

Coming back to convex records, the exchangeability of the data points
implies that the three cases listed in~(\ref{nr4}) and shown in figure~\ref{34}
occur with respective probabilities
\beq
q_1=q,\qquad
q_2=3q,\qquad
q_3=1-4q=1-P.
\eeq
We have therefore
\beq
\mean{N_4}=4(1-q)=4-P,\qquad
\mean{R_4}=4-q.
\label{nraves4}
\eeq
These mean values obey the identity~(\ref{nriden}), as should be.

In the asymptotic regime of most interest where $n$ becomes very large,
it is legitimate to use a continuum approximation,
so that~(\ref{nriden}) becomes\footnote{Here and throughout the following,
$x\approx y$ means that $x$ and $y$ are asymptotically equivalent
in the appropriate regime (here, $n\gg1$),
in the strong sense that $y/x$ converges to unity,
whereas the weaker form $x\sim y$
means that $y/x$ has much slower variations than $x$ or $y$ taken separately.}
\beq
\mean{N_n}\approx n\,\frac{\dd\mean{R_n}}{\dd n}\approx\frac{\dd\mean{R_n}}{\dd\ln n}.
\label{nrcont}
\eeq

The statistics of the number $N_n$ of vertices of the convex hull of $n$ iid points
in $d$-dimensional space, and especially in the plane,
has been the subject of a rather abundant mathematical literature
since the pioneering works by R\'enyi and Sulanke~\cite{RS63} and by Efron~\cite{E65}.
Most available rigorous results concern the mean number~$\mean{N_n}$ of vertices.
As it turns out, the behaviour of this quantity at large $n$
strongly depends on the underlying distribution of the data points.
The identity~(\ref{nriden}) enables us to predict in each case
the behaviour of the mean number~$\mean{R_n}$ of records.

Hereafter we consider four characteristic examples of iid data points in the plane,
namely uniform points in the square (section~\ref{square}),
uniform points in the disk (section~\ref{disk}),
Gaussian points (section~\ref{gauss}),
and points with an isotropic power-law distribution (section~\ref{levy}).

\subsection{Uniform points in the square}
\label{square}

We begin with a reminder on the more general situation of uniform points
in a convex polygon with $r$ sides and vertices $\R_i$.
The number $N_n$ of vertices of the convex hull of $n$ points
has been shown to obey a central limit theorem at large $n$,
i.e., to have an asymptotic Gaussian distribution,
whose mean and variance grow logarithmically with~$n$~\cite{G88}.

More precisely, the mean value of $N_n$ reads asymptotically~\cite{RS63}
\beq
\mean{N_n}\approx\frac{2r}{3}\left(\ln n+\frac{1}{r}\sum_{i=1}^r\ln\frac{\area_i}{\area}+\gamma\right),
\label{argon}
\eeq
where $\gamma$ is Euler's constant,
$\area$ is the total area of the polygon,
and $\area_i$ is the area of the triangle $\R_{i-1}\R_i\R_{i+1}$ formed by three consecutive vertices.
The area ratios entering the above formula are affine invariants.
For the regular $r$-gon of the Euclidean plane, this reads
\beq
\mean{N_n}\approx\frac{2r}{3}\left(\ln\left(\frac{4n}{r}\sin^2\frac{\pi}{r}\right)+\gamma\right).
\label{areg}
\eeq
When the number $r$ of sides becomes itself large,
the above result simplifies to
\beq
\mean{N_n}\approx\frac{2r}{3}\left(\ln\frac{4\pi^2n}{r^3}+\gamma\right).
\label{rlarge}
\eeq
This formula exhibits a crossover for $n\sim r^3$,
with $\mean{N_n}$ scaling as~(\ref{rlarge}) for $1\ll r^3\ll n$,
and as~(\ref{ndisk}) for $1\ll n\ll r^3$.

The variance of $N_n$ grows as~\cite{G88}
\beq
\var N_n\approx\frac{10r}{27}\,\ln n.
\label{vrgon}
\eeq
To our knowledge, no formula is known for the finite part of this logarithm.

For uniform points in a polygonal domain,
(\ref{argon}) and~(\ref{vrgon}) show that $\mean{N_n}$ and $\var N_n$
share the same logarithmic growth law in $n$.
The ratio
\beq
F_{N_n}=\frac{\var N_n}{\mean{N_n}},
\label{fano}
\eeq
known as the Fano factor of the distribution of $N_n$, goes to the limit
\beq
F_N=\frac{5}{9},
\label{fncarre}
\eeq
irrespective of the number $r$ of sides of the polygon.
The Fano factor~\cite{fano} is used
to characterise distributions of integers counting detected particles
and similar discrete events (see e.g.~\cite{TTT}).
Poisson distributions have a Fano factor $F=1$.
Distributions with~$F$ less than unity (resp.~larger than unity)
are referred to as sub-Poissonian (resp.~super-Poissonian).\footnote{The Mandel
parameter $Q=F-1$ is used in other areas of physics, including quantum optics~\cite{mandel}.}

We now turn to the specific case of uniform points in the unit square.
Combinatorial methods give access to some results for $N_n$ for finite $n$~\cite{B06,B09}.
When~$n$ is large, the asymptotic results~(\ref{areg}) and~(\ref{vrgon}) read
\beq
\mean{N_n}\approx A_1\ln n,\quad
\var N_n\approx A_2\ln n,\qquad
A_1=\frac{8}{3},\quad
A_2=\frac{40}{27}.
\label{avcarre}
\eeq
The identity~(\ref{nriden}) predicts that the mean value of $R_n$ grows as
\beq
\mean{R_n}\approx B_1(\ln n)^2,\qquad B_1=\frac{A_1}{2}=\frac{4}{3}.
\label{arcarre}
\eeq

\begin{figure}
\begin{center}
\includegraphics[angle=0,width=0.4\linewidth,clip=true]{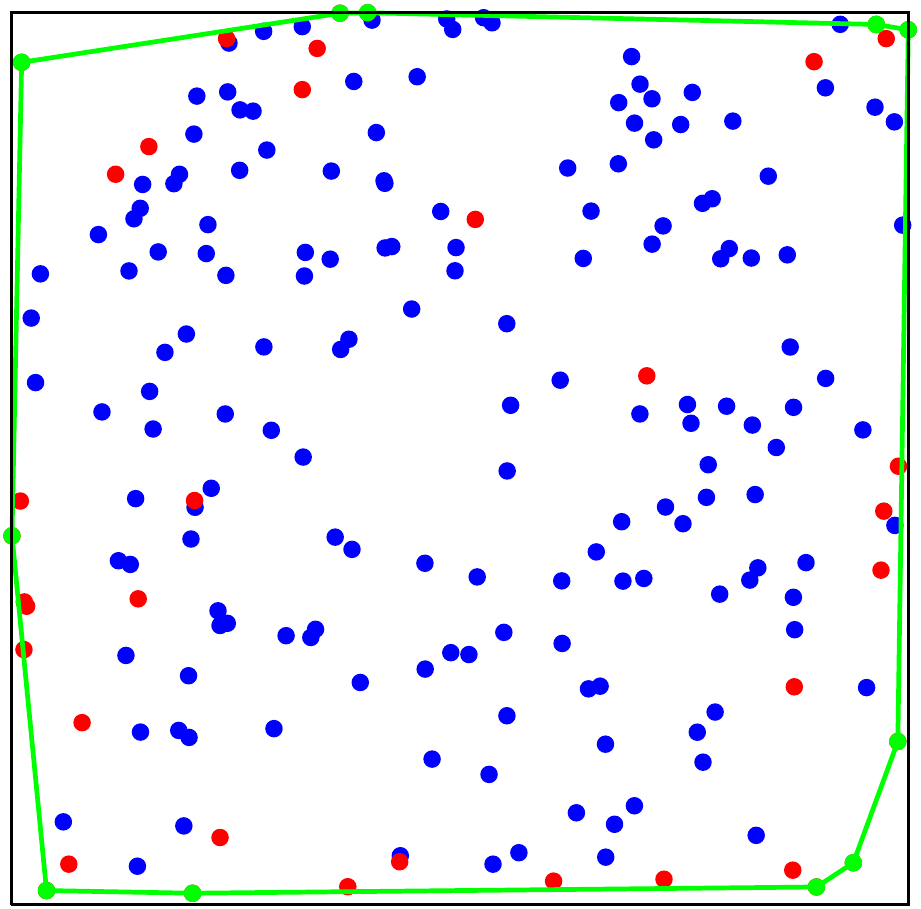}
\caption{\small
A sample of 200 uniform iid points in the unit square
such that $N=11$ and $R=38$.
Green polygon: convex hull of the dataset.
Green symbols: the 11 vertices of the convex hull.
Red symbols: the 27 other convex records.
Blue symbols: the 162 data points that are not convex records.}
\label{plotcarre}
\end{center}
\end{figure}

\begin{figure}
\begin{center}
\includegraphics[angle=0,width=0.6\linewidth,clip=true]{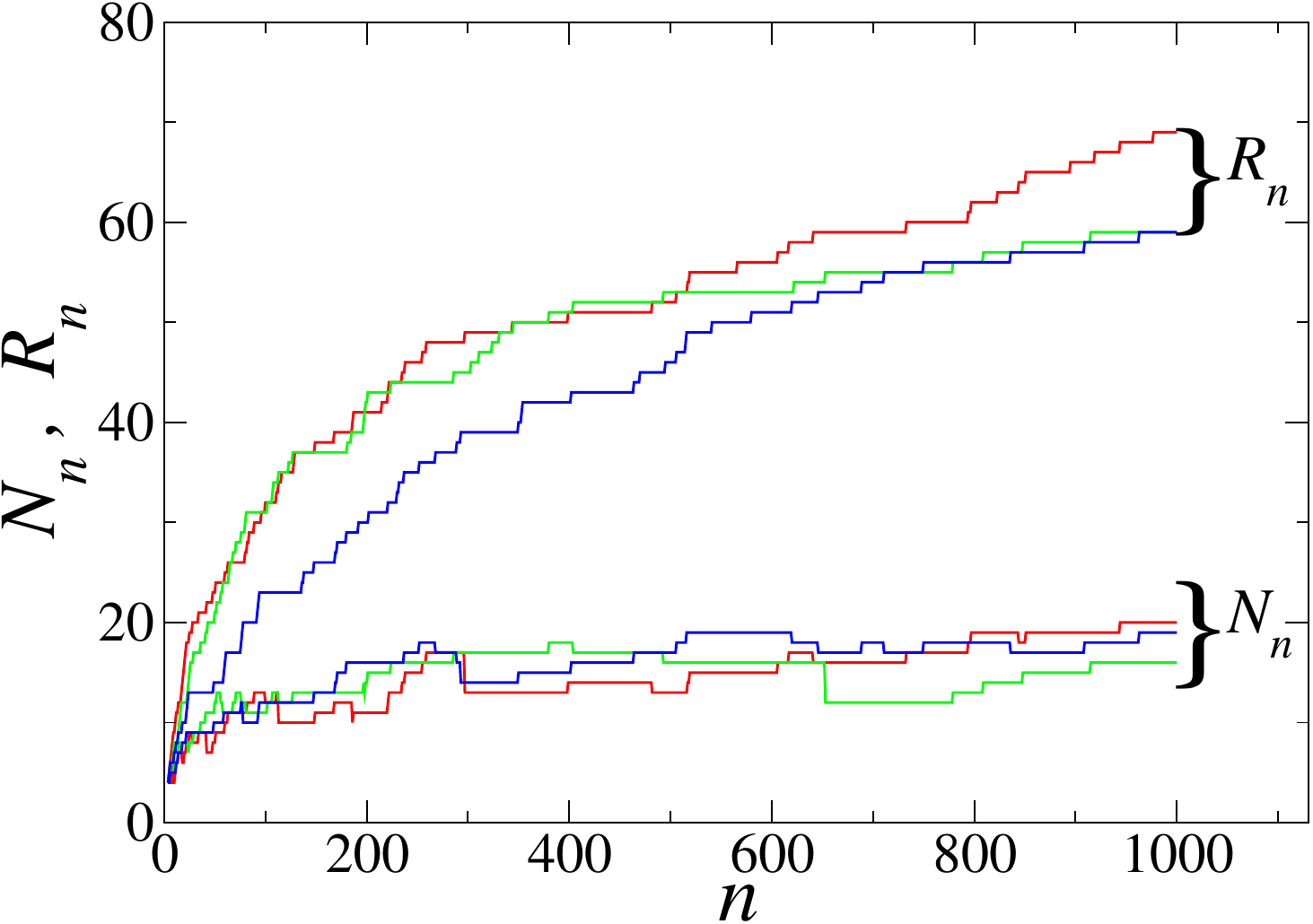}
\caption{\small
Numbers $N_n$ (lower tracks) and $R_n$ (upper tracks)
plotted against time~$n$ for three histories of 1,000 uniform data points in the unit square.
Each history is shown by a colour.}
\label{tracks}
\end{center}
\end{figure}

We have performed extensive numerical simulations
on the statistics of convex records.
For that purpose, we have developed a recursive algorithm constructing
the convex hull of a set of points in the plane that are added one by one.
As a first illustration,
we show in figure~\ref{plotcarre} the outcome of a simulation of 200 iid data points
in the unit square,
such that $N=11$ and $R=38$.
Figure~\ref{tracks} illustrates the evolution of $N_n$ (lower tracks) and~$R_n$ (upper tracks)
against time $n$ for three typical histories of 1,000 data points.
Each history is shown by a colour.
The numbers $N_n$ of vertices exhibit non-monotonic fluctuations as a function of $n$.
The numbers $R_n$ of records increase faster than~$N_n$,
and monotonically in time, as should be.

\begin{figure}
\begin{center}
\includegraphics[angle=0,width=0.6\linewidth,clip=true]{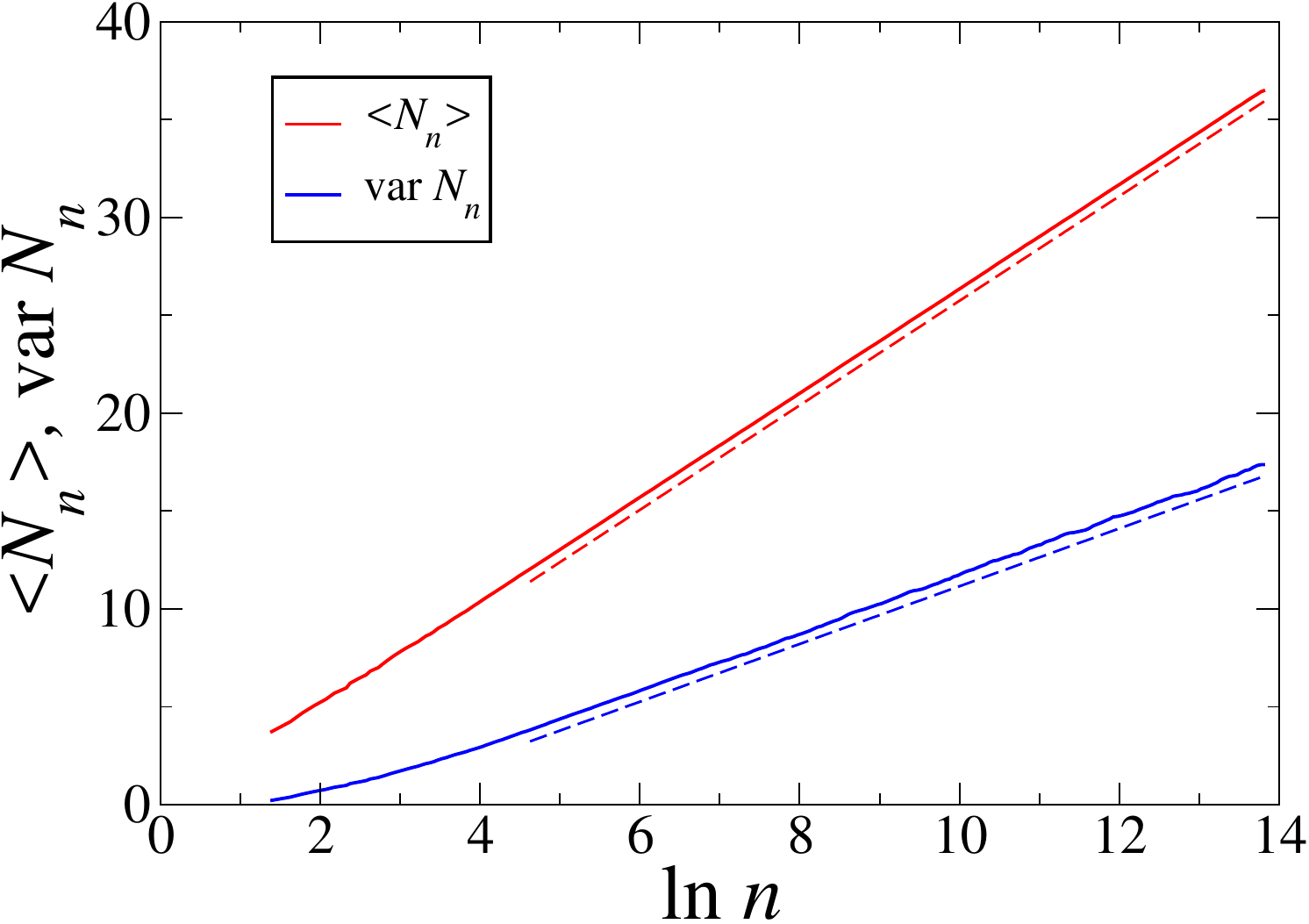}
\caption{\small
$\mean{N_n}$ and $\var{N_n}$ plotted against $\ln n$
for uniform points in the unit square.
Dashed lines show linear fits, slightly offset for greater clarity.}
\label{carre1}
\end{center}
\end{figure}

\begin{figure}
\begin{center}
\vskip 6pt
\includegraphics[angle=0,width=0.6\linewidth,clip=true]{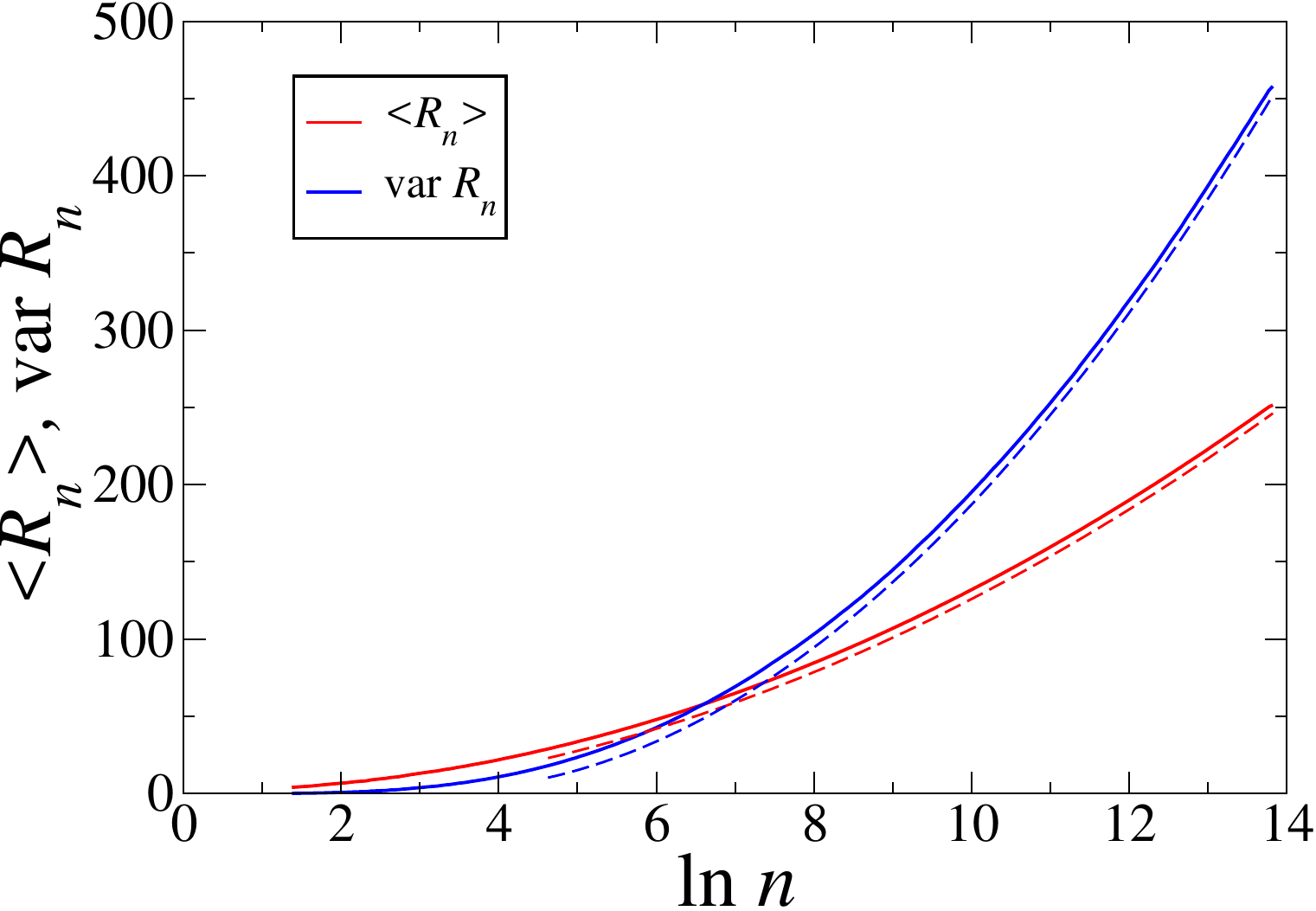}
\caption{\small
$\mean{R_n}$ and $\var{R_n}$ plotted against $\ln n$
for uniform points in the unit square.
Dashed curves show quadratic fits, slightly offset vertically for greater clarity.}
\label{carre2}
\end{center}
\end{figure}

To come to a quantitative study,
we focus our attention on the mean values and variances of $N_n$ and~$R_n$.
Figure~\ref{carre1} shows plots of $\mean{N_n}$ and $\var{N_n}$ against $\ln n$.
Here and throughout the following,
numerical data are gathered over~$10^5$ independent histories of $10^6$ data points each,
and only data for $n>100$ (sometimes $n>1,000$) are included in the analysis
of their asymptotic behaviour.
Dashed lines show linear fits to these data,
whose respective slopes~2.67 and 1.47
are to be compared with $A_1=8/3=2.666666\dots$ and $A_2=40/27=1.481481\dots$
(see~(\ref{avcarre})).
This very good agreement provides a solid validation of our numerical approach.
Figure~\ref{carre2} shows plots of $\mean{R_n}$ and $\var{R_n}$ against~$\ln n$.
Dashed curves show quadratic fits.
The coefficient of $(\ln n)^2$ for $\mean{R_n}$ reads $1.33$,
again in good agreement with the analytical prediction $B_1=4/3=1.333\dots$
(see~(\ref{arcarre})).
The coefficient of $(\ln n)^2$ for $\var R_n$ reads $3.96$,
yielding the growth law
\beq
\var R_n\approx B_2(\ln n)^2,\qquad B_2\approx3.96.
\label{vrcarre}
\eeq
This is the first amplitude for which there is no analytical prediction.
Equations~(\ref{arcarre}) and~(\ref{vrcarre}) yield the finite limit Fano factor
\beq
F_R=\frac{B_2}{B_1}=\frac{3B_2}{4}\approx2.97.
\label{frcarre}
\eeq

\subsection{Uniform points in the disk}
\label{disk}

We now consider uniform iid points in the unit disk.
Some combinatorial results on
the number $N_n$ of vertices of the convex hull of $n$ points in the disk
are available for finite~$n$~\cite{M17}.
For large $n$, $N_n$ obeys a central limit theorem,
i.e., it has an asymptotic Gaussian distribution~\cite{G88},
whose mean and variance grow as
\beq
\mean{N_n}\approx A_1\,n^{1/3},\qquad
\var N_n\approx A_2\,n^{1/3}.
\label{ndisk}
\eeq
These power laws are common to all finite convex domains of the plane with a smooth boundary.
In the case of the disk, the prefactors $A_1$ and $A_2$ are known exactly.
They read~\cite{RS63}
\beq
A_1=\left(\frac{2^7\pi^2}{3^4}\right)^{1/3}\Gamma(2/3)=3.383228\dots
\label{a1disk}
\eeq
and~\cite{FH04}
\beq
A_2=\left(\frac{2^7\pi^2}{3^{13}}\right)^{1/3}
\left(\frac{16\pi^2}{\Gamma(2/3)^2}-57\,\Gamma(2/3)\right)=0.826885\dots
\label{a2disk}
\eeq
The corresponding limit Fano factor is therefore
\beq
F_N=\frac{A_2}{A_1}=\frac{16\pi^2}{27\,\Gamma(2/3)^3}-\frac{19}{9}=0.244407\dots
\label{fncercle}
\eeq
The identity~(\ref{nriden}) predicts that the mean value of $R_n$ grows as
\beq
\mean{R_n}\approx B_1\,n^{1/3},\qquad B_1=3A_1=10.149686\dots
\label{ardisk}
\eeq

\begin{figure}
\begin{center}
\includegraphics[angle=0,width=0.6\linewidth,clip=true]{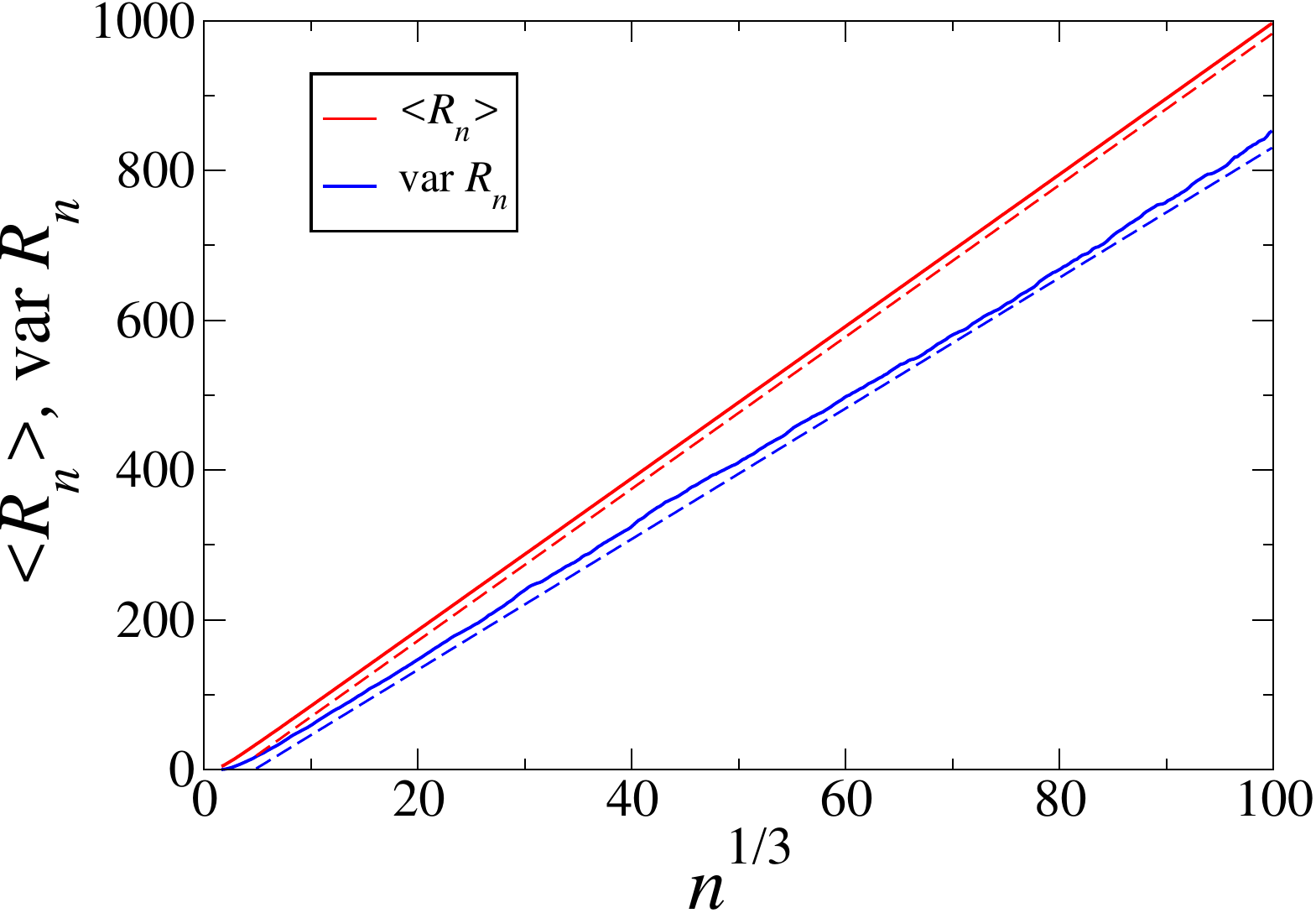}
\caption{\small
$\mean{R_n}$ and $\var{R_n}$ plotted against $n^{1/3}$
for uniform points in the unit disk.
Dashed lines show linear fits, slightly offset for greater clarity.}
\label{disk2}
\end{center}
\end{figure}

We have again measured the mean values and variances of $N_n$ and~$R_n$
by extensive numerical simulations.
Our data concerning $N_n$ (not shown)
are in very good agreement with~(\ref{a1disk}),~(\ref{a2disk}).
Figure~\ref{disk2} shows plots of $\mean{R_n}$ and $\var{R_n}$ against~$n^{1/3}$.
Dashed lines show linear fits to the data.
The slope for $\mean{R_n}$ reads~10.15,
in excellent agreement with the analytical prediction~(\ref{ardisk}).
The slope for $\var R_n$ reads 8.73,
implying the growth law
\beq
\var R_n\approx B_2\,n^{1/3},\qquad B_2\approx8.7.
\label{vrdisk}
\eeq
Equations~(\ref{ardisk}) and~(\ref{vrdisk}) yield the limit Fano factor
\beq
F_R=\frac{B_2}{B_1}\approx0.86.
\label{frcercle}
\eeq

\subsection{Gaussian points}
\label{gauss}

This section is devoted to iid Gaussian (or normal) points.
The invariance under the affine group
can be used to map an arbitrary Gaussian distribution
on a centred isotropic one such that $\mean{\abs{\x}^2}=1$.

Many papers in the mathematical literature deal with the convex hull
of two-dimensional Gaussian points~\cite{RS63,E65,C70,H94,H99}.
Exact formulas for the mean values of several quantities for finite $n$
have been derived in~\cite{E65}.
The expression for the mean number $N_n$ of vertices reads
\beq
\mean{N_n}=\sqrt{4\pi}\,J_n,
\label{gave}
\eeq
with
\beq
J_n=n(n-1)\int_{-\infty}^\infty F(x)^{n-2}f(x)^2\,\dd x\qquad(n\ge3),
\eeq
where
\beq
f(x)=\frac{\e^{-x^2/2}}{\sqrt{2\pi}},\qquad
F(x)=\frac{1}{2}\left(1+\erf\frac{x}{\sqrt2}\right)
\eeq
are the density and the distribution function of a Gaussian variable such that $\mean{x^2}=1$.
The integrals $J_n$ have been investigated in detail in~\cite{KL},
where they are denoted by~$A_n$.
Besides~(\ref{nr123}), the expressions
\beqa
&\mean{N_4}=\frac{6}{\pi}\arccos\left(-\frac{1}{3}\right)=3.649040\dots,
\label{gn4}
\\
&\mean{N_5}=\frac{5}{2\pi}\arccos\left(-\frac{23}{27}\right)=4.122601\dots
\eeqa
seem to exhaust the list of available closed-form results.
The asymptotic behaviour of~(\ref{gave}) reads
\beq
\mean{N_n}=(8\pi\ln n)^{1/2}
\left(1+\frac{\mu}{2\ln n}-\frac{\mu^2+2\mu+2+\pi^2/6}{8(\ln n)^2}+\cdots\right),
\label{nfull}
\eeq
with
\beq
\mu=\gamma-\frac{1}{2}\ln(4\pi\ln n),
\label{nudef}
\eeq
where $\gamma$ is Euler's constant.

The $(\ln n)^{1/2}$ growth law~(\ref{nfull}) is rather ubiquitous among data points
with isotropic distributions in the plane whose complementary radial distribution function
\beq
\prob(\abs{\x}>r)=F(r)=\exp(-\Phi(r))
\label{Fdef}
\eeq
decays rapidly as $r\to\infty$.
This situation has been investigated long ago by Carnal~\cite{C70},
resulting in the following parametric representation of the mean number $\mean{N_n}$ of vertices:
\beq
\ln n\approx\Phi(r),\qquad
\mean{N_n}\approx\left(4\pi r\Phi'(r)\right)^{1/2},
\label{carne}
\eeq
where the accent denotes a derivative.
The parameter $r$ has a simple interpretation:
in line with the theory of extreme-value statistics,
$r$ provides an estimate for the largest radius of the first $n$ data points.
For distributions falling off as a stretched or compressed exponential of the form
\beq
F(r)\sim\exp(-Ar^\alpha),
\eeq
(\ref{carne}) predicts
\beq
\mean{N_n}\approx(4\pi\alpha\ln n)^{1/2},
\eeq
where the exponent $\alpha$ only enters the prefactor of the $(\ln n)^{1/2}$ growth law.
In the limit situation of distributions decaying as a power law of the form
\beq
F(r)\approx\frac{c}{r^\theta},
\eeq
(\ref{carne}) predicts that the mean number of vertices saturates to the finite value
\beq
\mean{N_n}\approx\sqrt{4\pi\theta}.
\eeq
This is indeed the correct leading-order result
for power-law data points with a large exponent $\theta$ (see~(\ref{arac})).

Coming back to Gaussian data points,
the random number $N_n$ of vertices obeys a central limit theorem at large~$n$,
i.e., it has an asymptotic Gaussian distribution~\cite{H94,H99},
whose mean and variance grow as
\beq
\mean{N_n}\approx A_1(\ln n)^{1/2},\qquad
\var N_n\approx A_2(\ln n)^{1/2}.
\label{ngau}
\eeq
The prefactor
(see~(\ref{nfull}))
\beq
A_1=\sqrt{8\pi}=5.013256\dots
\label{angau}
\eeq
has been long known~\cite{RS63}.
No formula for $A_2$ seems to be known to date (see~\cite{H94}).

The identity~(\ref{nriden}) predicts that the mean value of $R_n$ grows as
\beq
\mean{R_n}\approx B_1(\ln n)^{3/2},\qquad
B_1=\frac{2A_1}{3}=\frac{4\sqrt{2\pi}}{3}=3.342171\dots
\label{argau}
\eeq
In analogy with previous cases, we anticipate that the variance of $R_n$ also scales as
\beq
\var R_n\approx B_2(\ln n)^{3/2}.
\label{b2gau}
\eeq

\begin{figure}
\begin{center}
\includegraphics[angle=0,width=0.6\linewidth,clip=true]{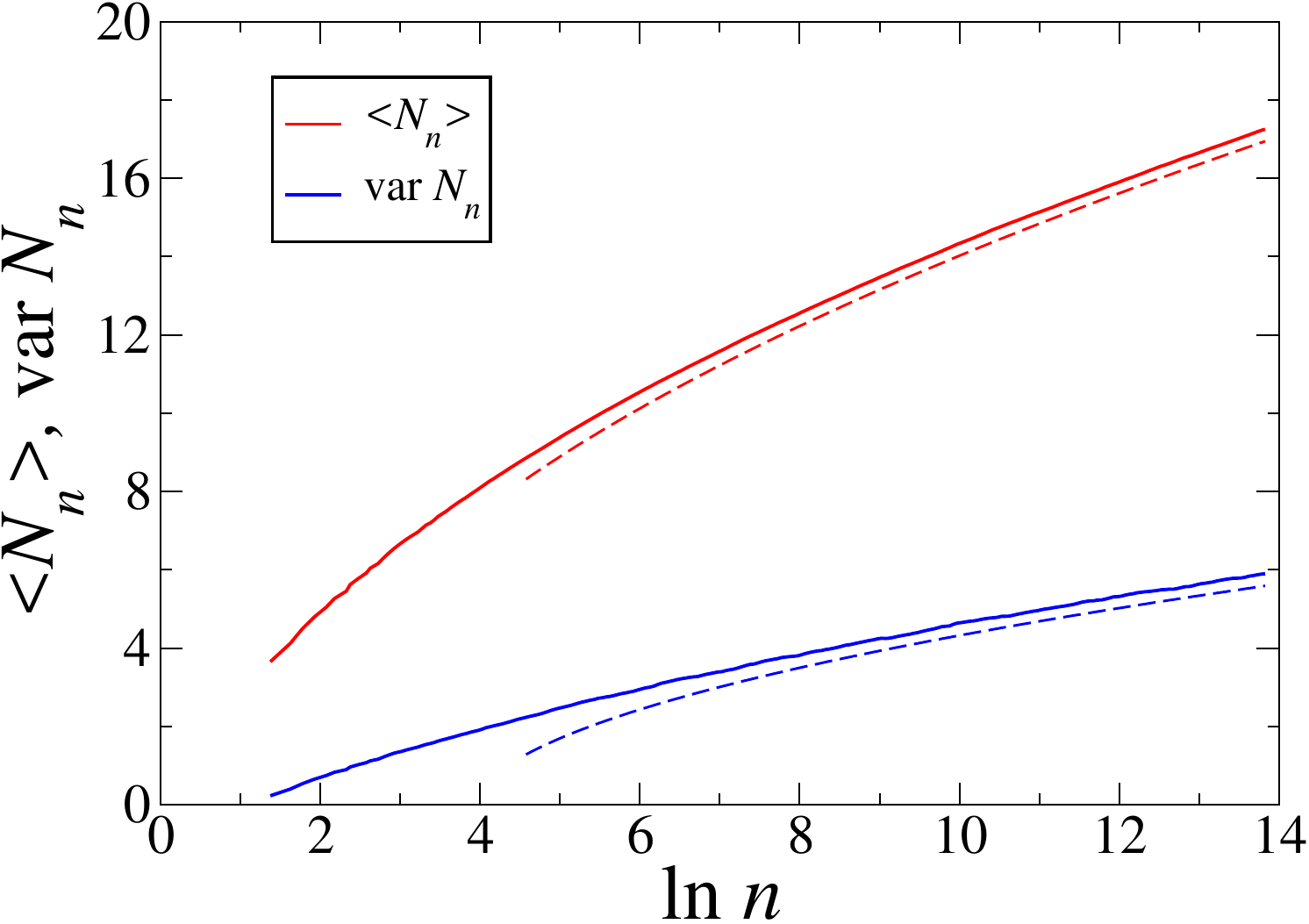}
\caption{\small
$\mean{N_n}$ and $\var{N_n}$ plotted against $\ln n$
for Gaussian points.
Dashed curves show the non-linear fits described in the text,
slightly offset vertically for greater clarity.}
\label{gauss1}
\end{center}
\end{figure}

\begin{figure}
\begin{center}
\includegraphics[angle=0,width=0.6\linewidth,clip=true]{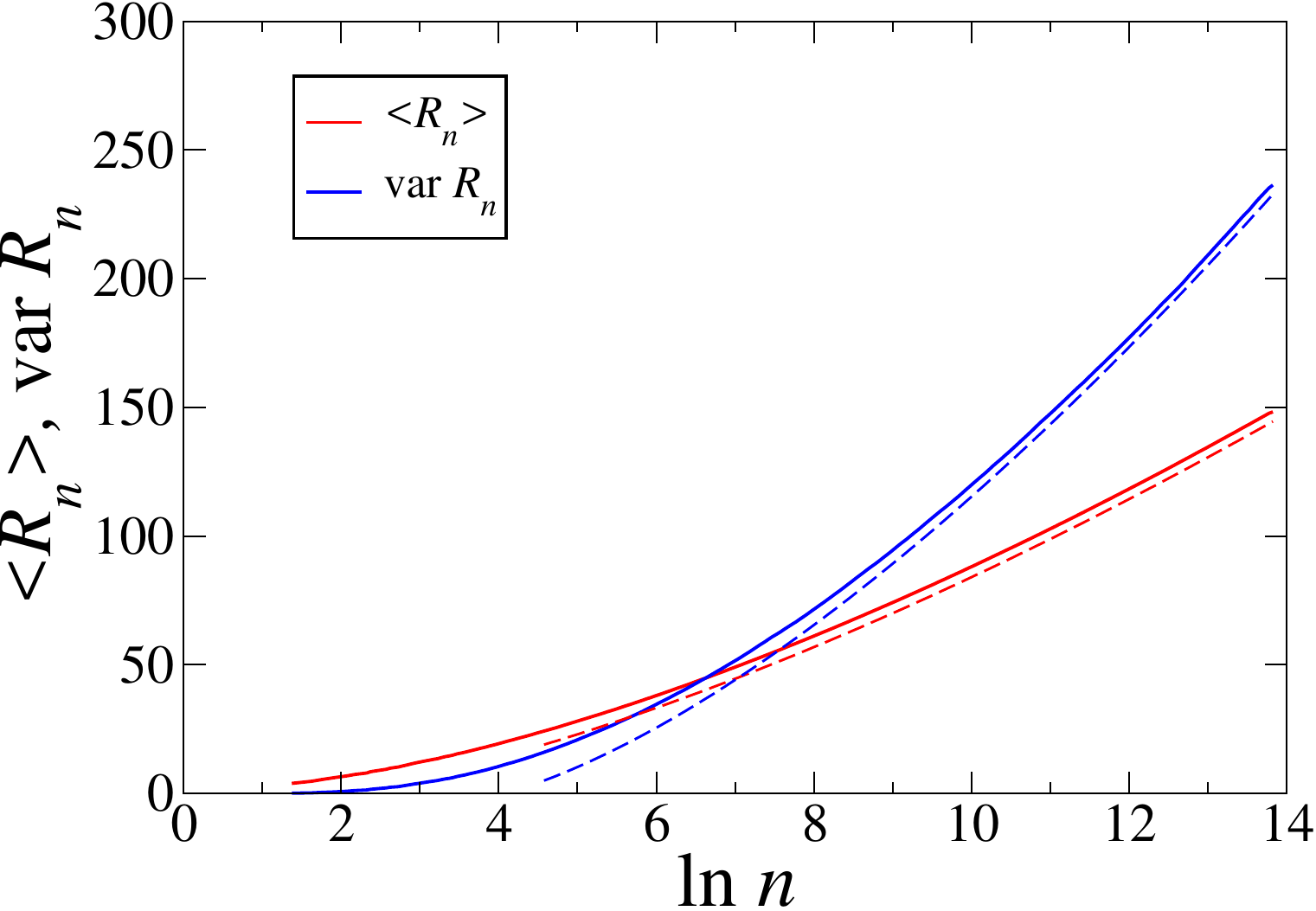}
\caption{\small
$\mean{R_n}$ and $\var{R_n}$ plotted against $\ln n$
for Gaussian points.
Dashed curves show the non-linear fits described in the text,
slightly offset vertically for greater clarity.}
\label{gauss2}
\end{center}
\end{figure}

We have measured the mean values and variances of $N_n$ and~$R_n$
by extensive numerical simulations.
Figure~\ref{gauss1} shows plots of $\mean{N_n}$ and $\var{N_n}$ against~$\ln n$.
Dashed curves show non-linear fits of the form $y=a(\ln n+b)^{1/2}$.
The complexity of the sequence of subleading corrections entering~(\ref{nfull})
has deterred us from using more sophisticated fits.
In the case of $\mean{N_n}$,
the fit parameter $a=4.911$, to be identified with $A_1$,
is in good agreement with the prediction~(\ref{angau})
(2 percent relative difference),
especially in view of the above.
In the case of $\var N_n$, the fit parameter $a=1.865$ yields
\beq
A_2\approx1.86,
\eeq
with an expected relative accuracy in the range of a few percent.
Figure~\ref{gauss2} shows plots of $\mean{R_n}$ and $\var{R_n}$ against $\ln n$.
Dashed curves show non-linear fits of the form $y=a(\ln n+b)^{3/2}$.
In the case of $\mean{R_n}$,
the parameter $a=3.17$, to be identified with~$B_1$,
is in reasonably good agreement with the prediction~(\ref{argau})
(5 percent relative difference).
In the case of $\var R_n$, the parameter $a=7.03$ yields
\beq
B_2\approx7.0,
\eeq
again with an expected relative accuracy of a few percent.

The above results translate to the limit Fano factors
\beq
F_N=\frac{A_2}{A_1}\approx0.37,\qquad
F_R=\frac{B_2}{B_1}\approx2.10.
\label{fgauss}
\eeq

\subsection{Points with isotropic power-law distributions}
\label{levy}

Our last example concerns iid data points $\x_n$ with an isotropic distribution
whose complementary radial distribution function (see~(\ref{Fdef}))
falls off as a power law at large distances, namely
\beq
F(r)\approx\frac{c}{r^\th},
\label{fth}
\eeq
with an arbitrary exponent $\th>0$.

A few works in the mathematical literature deal with the convex hull
of such iid random points~\cite{C70,H99,A91}.
At variance with previous examples,
the distribution of the number $N_n$ of its vertices now reaches a finite limit,
\beq
p_k(\th)=\lim_{n\to\infty}\prob(N_n=k)\qquad(k\ge3),
\eeq
as the number $n$ of points becomes infinitely large.
The above limit distribution is universal,
in the sense that it only depends on the exponent $\th$.
For generic values of~$\th$, only the first moment
\beq
A_1(\th)=\lim_{n\to\infty}\mean{N_n}=\sum_{k\ge3}k p_k(\th)
\label{anlevy}
\eeq
is known explicitly and reads
\beq
A_1(\th)=4\sqrt{\pi}\,
\frac{\Gamma^2(\half\th+1)\,\Gamma(\th+\half)}{\Gamma^2(\half\th+\half)\,\Gamma(\th+1)}.
\label{a1}
\eeq
This expression starts from $A_1(0)=4$ (see~(\ref{a0})),
and grows at large $\th$ as
\beq
A_1(\th)=(4\pi\th)^{1/2}\left(1+\frac{3}{8\th}+\frac{9}{128\th^2}+\cdots\right).
\label{arac}
\eeq
We introduce for further reference the notation for the corresponding variance:
\beq
A_2(\th)=\lim_{n\to\infty}\var N_n=\sum_{k\ge3} k^2p_k(\th)-A_1(\th)^2.
\label{vnlevy}
\eeq
We mention for completeness that the problem simplifies
in the $\th\to0$ limit~\cite{A91},
where the full distribution $p_k(0)$ has been obtained explicitly:
\beq
p_k(0)=2^{k-3}
\Biggl(2\,\frac{(\ln 2)^{k-2}}{(k-2)!}-2+\sum_{j=0}^{k-3}\frac{(\ln 2)^j}{j!}\Biggr).
\eeq
The corresponding generating function,
\beq
G(z)=\lim_{n\to\infty}\mean{z^{N_n}}=\sum_{k\ge3}p_k(0)z^k=\frac{z^2}{1-2z}\,((1-z)2^{2z}-1),
\eeq
yields in particular
\beq
A_1(0)=4,\qquad A_2(0)=16\ln 2-10=1.090354\dots
\label{a0}
\eeq
The identity~(\ref{nriden}) predicts that the mean value of $R_n$ grows as
\beq
\mean{R_n}\approx B_1(\th)\,\ln n,\qquad B_1(\th)=A_1(\th).
\label{arlevy}
\eeq
In analogy with previous cases, we anticipate that the variance of $R_n$ scales as
\beq
\var R_n\approx B_2(\th)\,\ln n.
\label{vrlevy}
\eeq

\begin{figure}[!ht]
\begin{center}
\includegraphics[angle=0,width=0.6\linewidth,clip=true]{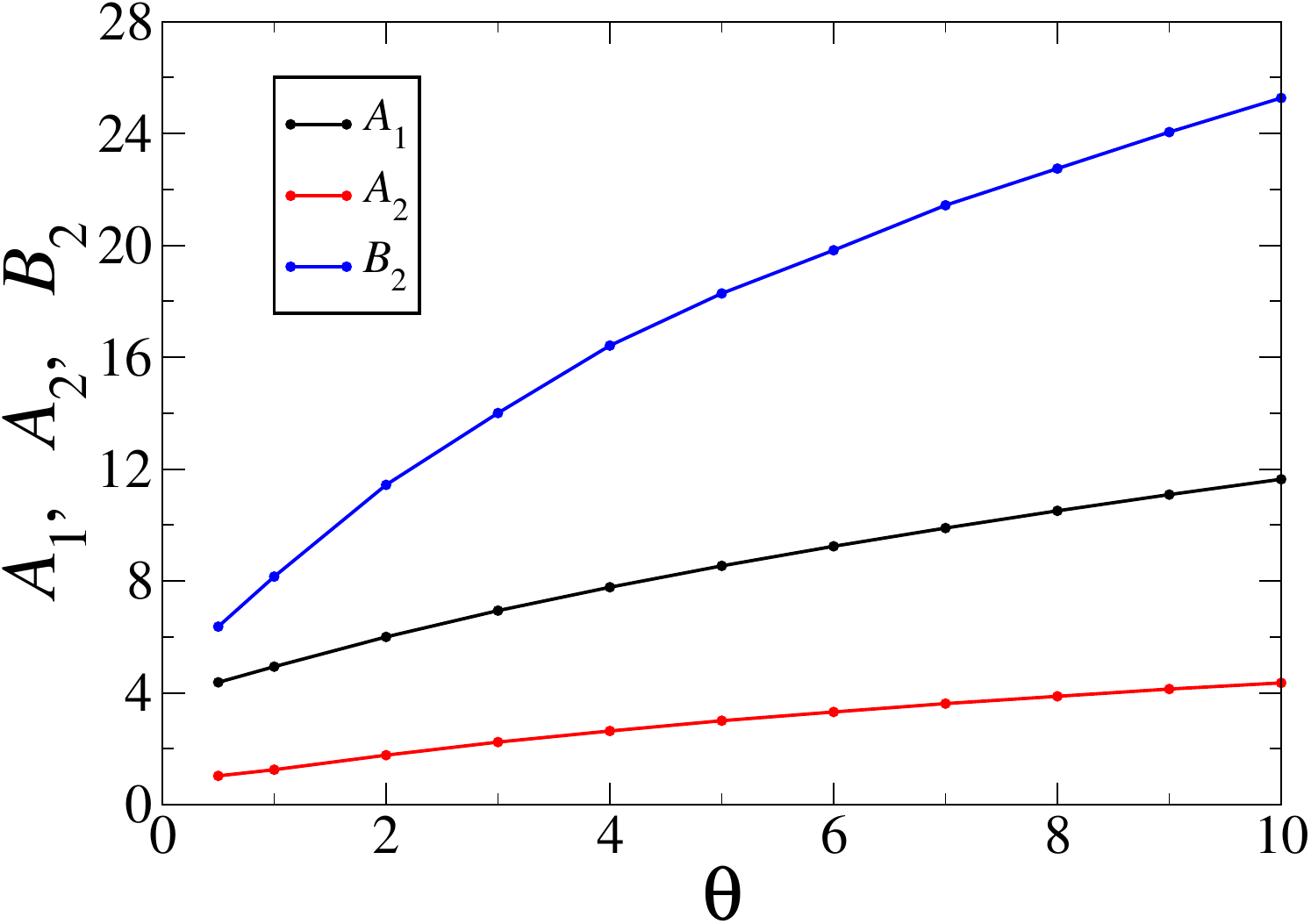}
\caption{\small
Amplitudes $A_1(\th)$, $A_2(\th)$ and $B_2(\th)$
characterising convex hulls and convex records of data points with isotropic power-law distributions,
plotted against the exponent $\th$.
Black: $A_1(\th)$ entering~(\ref{anlevy}) and~(\ref{arlevy}), and given by~(\ref{a1}).
Red: numerical values of $A_2(\th)$ entering~(\ref{vnlevy}).
Blue: numerical values of $B_2(\th)$ entering~(\ref{vrlevy}).}
\label{laplot}
\end{center}
\end{figure}

\begin{figure}[!ht]
\begin{center}
\includegraphics[angle=0,width=0.6\linewidth,clip=true]{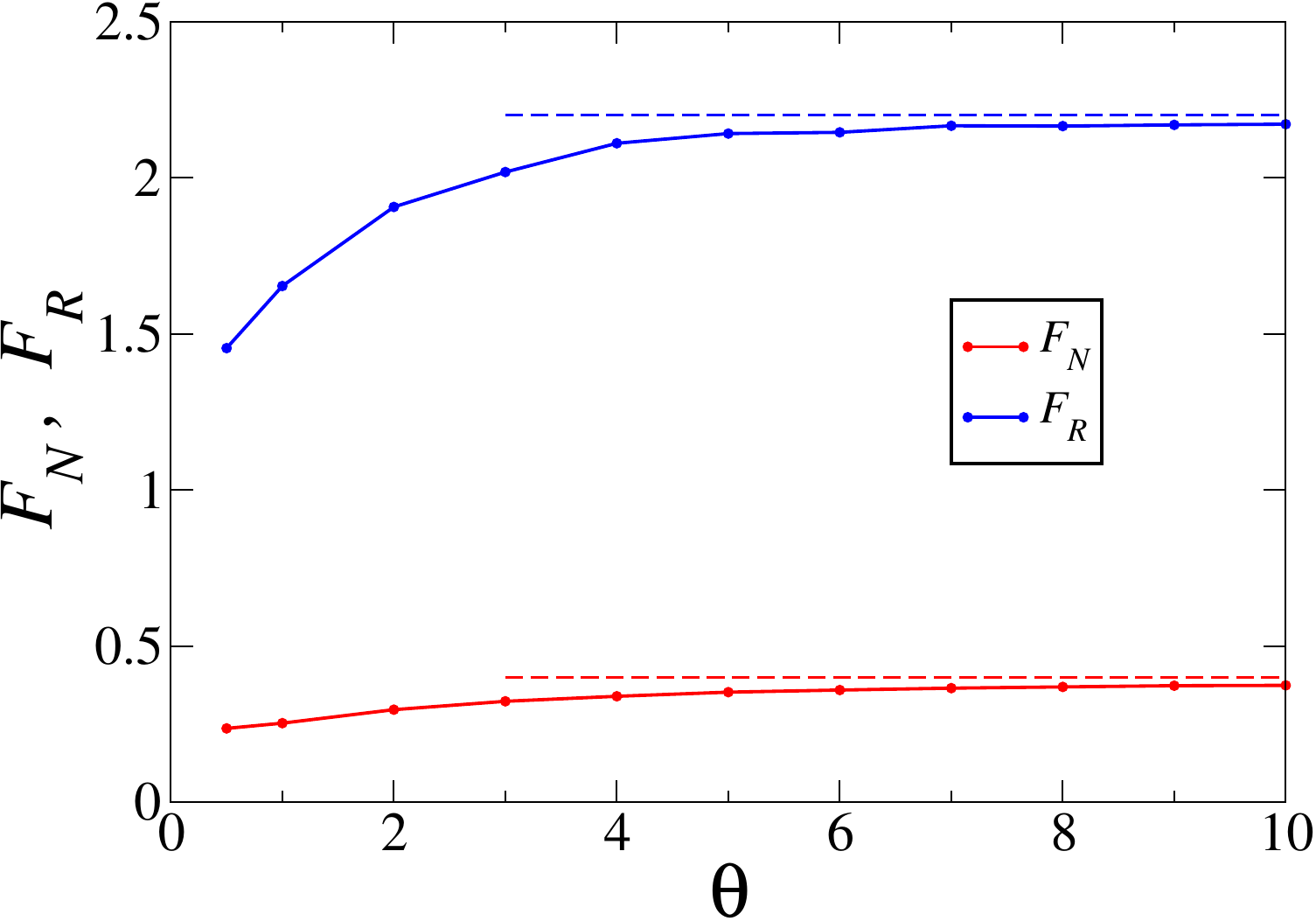}
\caption{\small
Fano factors defined in~(\ref{lfano}), plotted against the exponent $\th$.
Red: $F_N(\th)$.
Blue: $F_R(\th)$.
Horizontal dashed lines: estimated limits~(\ref{llims}).}
\label{lfplot}
\end{center}
\end{figure}

We have run extensive numerical simulations for exponents $\th$ ranging from $1/2$ to~10.
Data points with isotropic power-law distributions
were generated by using the non-linear mapping
\beq
\x=\frac{\y}{(1-\abs{\y})^{1/\th}}.
\eeq
If $\y$ is uniformly distributed in the unit disk,
the distribution of $\x$ is isotropic and obeys~(\ref{fth})
with $c=2$ and an arbitrary exponent $\th>0$.
We have again measured the mean values and variances of $N_n$ and~$R_n$.
Our data corroborate the above picture.
The measured $\mean{N_n}$ and $\var N_n$
go to well-defined limits $A_1(\th)$ and $A_2(\th)$, shown in figure~\ref{laplot}.
The observed values of $A_1(\th)$ are in very good agreement with the analytical result~(\ref{a1}).
The measured $\mean{R_n}$ and $\var R_n$ are found to follow
the logarithmic growth laws~(\ref{arlevy}) and~(\ref{vrlevy}).
The numerical values of $B_1(\th)$ are in very good agreement with the prediction~(\ref{arlevy}).
The values of $B_2(\th)$ are also shown in figure~\ref{laplot}.
This figure strongly suggests that the three plotted quantities share the same
square-root law at large $\th$ (see~(\ref{arac})).
This observation is corroborated by figure~\ref{lfplot}, showing the Fano factors
\beq
F_N(\th)=\frac{A_2(\th)}{A_1(\th)},\qquad
F_R(\th)=\frac{B_2(\th)}{B_1(\th)}.
\label{lfano}
\eeq
These quantities are found to converge to the limits
\beq
F_N(\infty)\approx0.40,\qquad
F_R(\infty)\approx2.20.
\label{llims}
\eeq
These values are rather close to those corresponding to Gaussian points (see~(\ref{fgauss})).

\subsection{A summary on mean values and variances}
\label{sum}

Our investigation of convex records
for four characteristic examples of iid points in the plane
(uniform points in the square and in the disk,
isotropic Gaussian and power-law points)
allows us to draw the following conclusions.
Consider first the number~$N_n$ of vertices of the convex hull of the first $n$ points.
In the first three examples recalled above,
$\mean{N_n}$ and $\var N_n$ grow proportionally to each other,
resulting in a finite limit Fano factor $F_N$.
For these examples,
and many other cases studied in the mathematical literature,
$N_n$ is known to obey a central limit theorem,
with mean values and variances growing at the same pace, as recalled above.
Note however that a counterexample `whose support is quite a complicated geometric object'
has been constructed~\cite{M00},
for which the relative fluctuations of $N_n$ around $\mean{N_n}$
do not shrink to zero as $n$ becomes large.
Our extensive numerical simulations have demonstrated
that the number $R_n$ of convex records behaves quite similarly,
in that $\mean{R_n}$ and $\var R_n$ also grow proportionally to each other,
also resulting in a finite Fano factor $F_R$.
We recall that the mean values $\mean{N_n}$ and $\mean{R_n}$
are related to each other by the identity~(\ref{nriden}).

\begin{figure}[!ht]
\begin{center}
\includegraphics[angle=0,width=0.6\linewidth,clip=true]{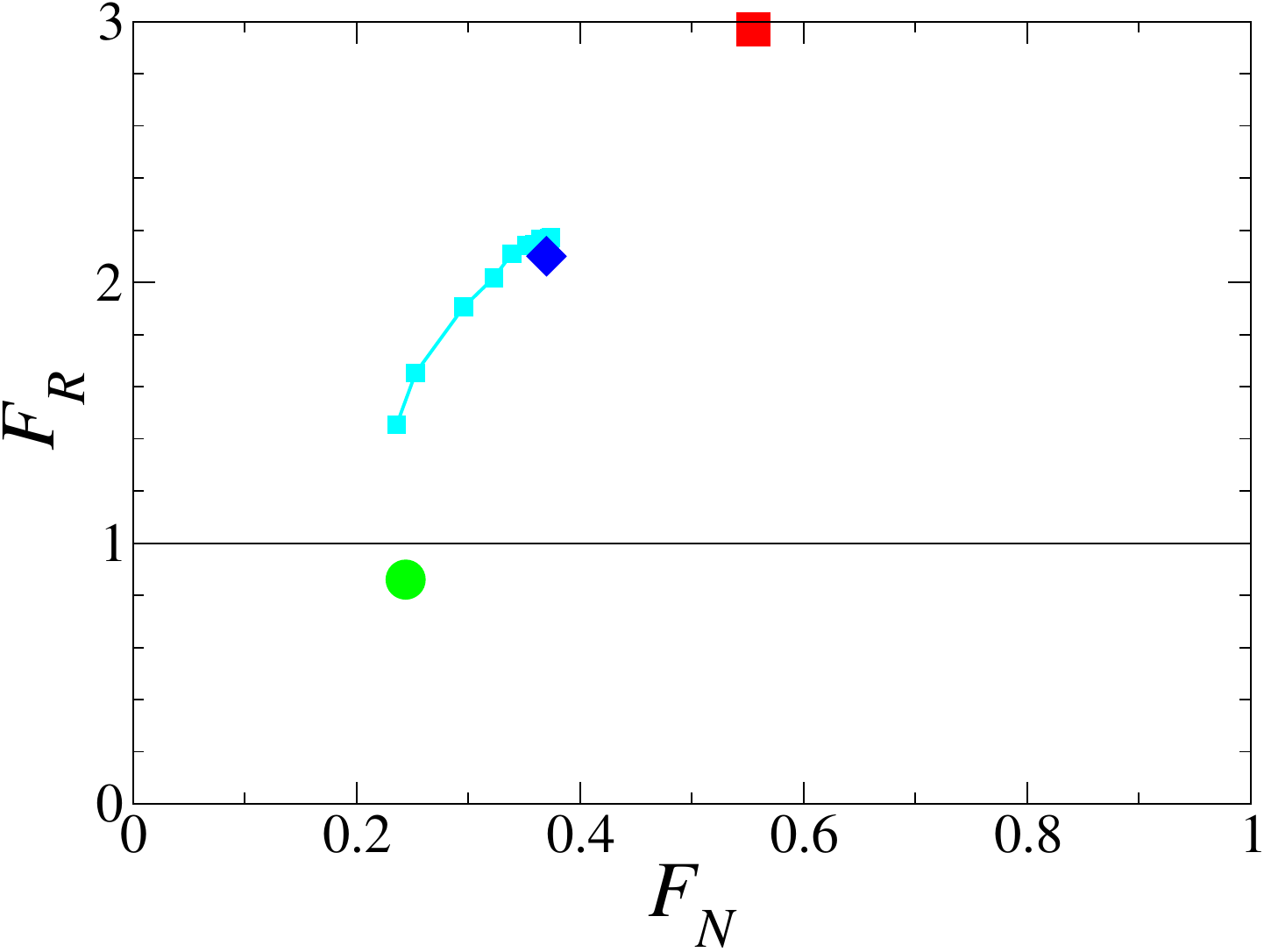}
\caption{\small
Scatter plot of the limit Fano factors $F_N$ and $F_R$
obtained for the four examples of iid points in the plane considered in this work.
Red square: uniform points in the square (section~\ref{square}).
Green disk: uniform points in the disk (section~\ref{disk}).
Blue diamond: Gaussian points (section~\ref{gauss}).
Cyan symbols joined by a line: Isotropic power-law points (section~\ref{levy}).}
\label{plotfanos}
\end{center}
\end{figure}

Table~\ref{iids} summarises the asymptotic behaviour of the mean values and variances
of~$N_n$ and $R_n$.
Amplitudes whose analytical expression was known exactly
are marked by stars.
Figure~\ref{plotfanos} shows a scatter plot of all Fano factors thus obtained.
Among these numbers, only the first two values of $F_N$ are known exactly
(see~(\ref{fncarre}),~(\ref{fncercle})).
In all cases,~$F_N$ is smaller than unity,
so that the distribution of~$N_n$ is asymptotically sub-Poissonian.
In all cases but the disk, $F_R$ is larger than unity,
so that the distribution of~$R_n$ is super-Poissonian.
The case of uniform points in the disk is somehow an outlier in two respects:
$\mean{N_n}$ and $\mean{R_n}$ grow as a power law, and the distribution of~$R_n$ is sub-Poissonian.
We recall that the distribution of $R_n$ is asymptotically Poissonian
in the classical case of univariate records (see~\ref{app}).

\begin{table}
\begin{center}
$$
\begin{array}{|l|c|c|c|c|c|}
\hline
\hbox{Distribution} & \hbox{section} & \mean{N_n} & \var{N_n} &
\mean{R_n} & \var{R_n} \\
\hline
\hbox{uniform (square)} & \hbox{(\ref{square})} & A_1^\star \ln n \hfill & A_2^\star \ln n \hfill &
B_1^\star(\ln n)^2 \hfill & B_2(\ln n)^2 \hfill \\
\hbox{uniform (disk)} & \hbox{(\ref{disk})} & A_1^\star\, n^{1/3} \hfill & A_2^\star\, n^{1/3} \hfill &
B_1^\star\, n^{1/3} \hfill & B_2\, n^{1/3} \hfill \\
\hbox{Gaussian} & \hbox{(\ref{gauss})} & A_1^\star\, (\ln n)^{1/2} \hfill & A_2\, (\ln n)^{1/2} \hfill &
B_1^\star\, (\ln n)^{3/2} \hfill & B_2\, (\ln n)^{3/2} \hfill \\
\hbox{power-law} & \hbox{(\ref{levy})} & A_1^\star(\th) \hfill & A_2(\th) \hfill &
B_1^\star(\th) \ln n \hfill & B_2(\th) \ln n \hfill \\
\hline
\end{array}
$$
\caption{Asymptotic behaviour of the mean values and variances of $N_n$ and $R_n$
for the four examples of iid data points studied in this work.
Amplitudes with a star in superscript are known analytically.
All other amplitudes are determined numerically.}
\label{iids}
\end{center}
\end{table}

\subsection{Extremal probabilities}
\label{ext}

The full distributions of $N_n$ and $R_n$ have many features of potential interest,
besides their mean values and variances investigated so far.
Hereafter we focus our attention on the extremal values of these random numbers,
namely 3 and $n$ (see~(\ref{nrtri}) and figure~\ref{NR}).

The probability that the number of records takes its minimal value $R_n=3$ reads~\cite{K95}
\beq
\prob(R_n=3)=\mean{A(\x_1,\x_2,\x_3)^{n-3}}\qquad(n>3),
\eeq
with the notation used in~(\ref{qdef}).
The meaning of this general result is clear:
$R_n=3$ holds for datasets of $n$ points where all subsequent $n-3$ points
fall inside the triangle formed by the first three ones.
Along this line of thought,
using again the exchangeability of the data points,
$N_n=3$ corresponds to datasets of $n$ points where the remaining $n-3$ points
fall inside the triangle formed by any three different points.
This observation translates to the identity
\beq
\prob(N_n=3)={n\choose3}\,\prob(R_n=3).
\label{priden}
\eeq
The event $R_n=3$ corresponds to the bottom left-hand corner of figure~\ref{NR},
whereas $N_n=3$ corresponds to the entire leftmost column.
The corresponding extremal probabilities only differ
by a simple combinatorial factor.
Equations~(\ref{nriden}) and~(\ref{priden})
are the only two general identities we have found for convex records of bivariate iid data.

The asymptotic behaviour of the extremal probability $\prob(N_n=3)$
depends on the underlying distribution of points.
For isotropic power-law points,
$\prob(N_n=3)$ goes to the universal limit $p_3(\th)$.
For uniform points inside a convex domain,
this probability generically falls off exponentially fast, as
\beq
\prob(N_n=3)\sim A_\star^n,
\label{astarexp}
\eeq
where $A_\star$ is the `probability content' of the largest triangle inscribed in the domain,
i.e., the fraction of the total area enclosed by this largest triangle.
We have
\beqa
&A_\star({\rm triangle})=1,\qquad
A_\star({\rm square})=\frac{1}{2},
\nonumber\\
&A_\star({\rm disk})=\frac{3\sqrt{3}}{4\pi}=0.413496\dots
\eeqa
In the case of a triangular domain, $A_\star=1$ prevents the exponential decay of~(\ref{astarexp}).
The extremal probabilities obey the power law $\prob(N_n=3)\approx8/n^3$~\cite{reed,alagar},
and so $\prob(R_n=3)\approx48/n^6$.

The probability that the number of vertices takes its maximal value $N_n=n$
has been investigated for uniform points in several domains.
Exact combinatorial results are available for the triangle and the square~\cite{V95,V96},
whereas an asymptotic analysis has been performed in the case of the disk~\cite{HCS08}.
This extremal probability is found to fall off super-exponentially, as
\beq
\prob(N_n=n)\sim\frac{B^n}{(n!)^2},
\label{fac2}
\eeq
with
\beqa
&B({\rm triangle})=\frac{27}{2}=13.5,\qquad
B({\rm square})=16,
\nonumber\\
&B({\rm disk})=2\pi^2=19.739208\dots
\eeqa

It is quite plausible that the exponential law~(\ref{astarexp})
and the $1/(n!)^2$ law~(\ref{fac2}) hold for a much larger class
of distributions of points.

\begin{figure}[!ht]
\begin{center}
\includegraphics[angle=0,width=0.6\linewidth,clip=true]{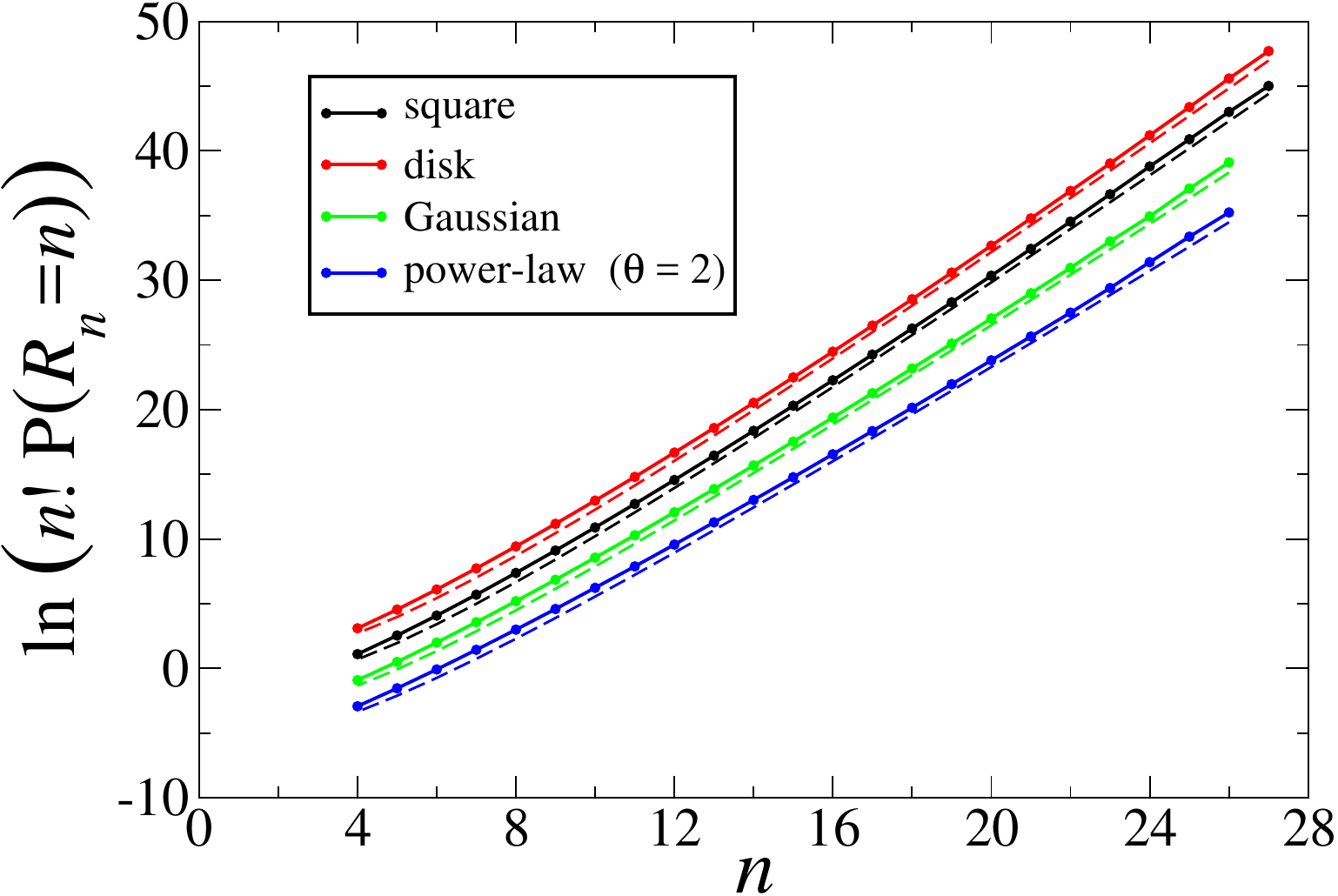}
\caption{\small
Logarithmic plot of the extremal probability $\prob(R_n=n)$, multiplied by $n!$, against $n$
for the above four characteristic examples of iid points in the plane.
Black: uniform points in the square.
Red: uniform points in the disk.
Green: Gaussian points.
Blue: power-law points with $\th=2$.
Dashed curves: fits $y=an+b\ln n+c$, so that $C=\e^a$.
Curves are slightly offset vertically from one another for greater clarity.}
\label{pex}
\end{center}
\end{figure}

The probability $\prob(R_n=n)$
that the number of records takes its maximal value $R_n=n$
is expected to fall off less rapidly than $\prob(N_n=n)$.
The event $N_n=n$ indeed corresponds to the top right-hand corner of figure~\ref{NR},
whereas $R_n=n$ corresponds to the entire top row.
We have run numerical simulations to measure $\prob(R_n=n)$
for the above four characteristic examples of iid points in the plane.
Figure~\ref{pex} shows logarithmic plots of the product $n!\,\prob(R_n=n)$ against time $n$.
The fits to the data (see caption) convincingly suggest the behaviour
\beq
\prob(R_n=n)\sim\frac{C^n}{n!},
\eeq
where the constant $C$ has a weak dependence on the underlying distribution of points:
\beqa
&C({\rm square})\approx9.7,\qquad
C({\rm disk})\approx10.2,
\nonumber\\
&C({\rm Gaussian})\approx8.5,\qquad
C(\hbox{power-law, $\th=2$})\approx7.5.
\eeqa

A plausible justification for the $1/n!$ decay of $\prob(R_n=n)$
is that the event $R_n=n$ corresponds to histories where every new data point
is outside the convex hull of all previous ones.
Loosely speaking, data point are further and further away from the origin.
In this regard, they resemble univariate records, for which $\prob(R_n=n)=1/n!$ exactly
(see~(\ref{unipex})).

\section{Random walks}
\label{rw}

We now consider convex records for data points generated by planar random walks.
The points~$\x_n$ are the successive positions of a random walker launched at the origin:
\beq
\x_n=\x_{n-1}+\del_n
\qquad(\x_0=\0).
\eeq
The increments $\del_n$ are iid two-dimensional random vectors,
drawn from some fixed distribution.
After $n$ time steps, the dataset consists of $n+1$ points $\x_0,\dots,\x_n$.

In this work we consider in parallel four examples of planar random walks:
the Pearson walk~\cite{Pearson},
where the increments~$\del_n$ are uniformly distributed over the unit circle,
and the P\'olya walk~\cite{Polya} on three lattices (triangular, square, hexagonal),
where the $\del_n$ take a finite number $z$ of discrete values,
equal to the coordination number of the lattice
(respectively 6, 4 and~3).
In all these models we have $\abs{\del_n}=1$, and so
\beq
\mean{\abs{\x_n}^2}=n.
\label{xdiff}
\eeq

Many rigorous results on the convex hulls of planar random walks and Brownian curves
have been derived, concerning in particular the mean value
of their area, perimeter length and number of vertices~\cite{SW61,B61,T80}
(see~\cite{MCR} for a synthetic review).
More recent developments on the combinatorics of random walks
in two and higher dimensions
also address properties of their convex hulls~\cite{S02,K17}.

Various types of records may be attached to planar random walks.
Three classes of such records, namely diagonal, simultaneous and radial ones,
have been investigated recently~\cite{usplanar}.
The mean numbers of these records grow as universal powers of time,
with respective exponents 1/4, 1/3 and 1/2.
Their full asymptotic distributions have also been determined.

Hereafter we consider convex records,
denoting by $N_n$ the number of vertices
of the convex hull of the walk at time $n$,
and by $R_n$ the number of convex records up to time~$n$.
We recall that $\x_n$ is not counted as a convex record
if it falls exactly on the boundary of the convex hull $\C(\x_0,\dots,\x_{n-1})$.
This event occurs with non-zero probability in the case of lattice walks.
Furthermore, to keep our numerical algorithm unchanged,
we have imposed the constraint that the first two steps $\del_1$ and $\del_2$
of lattice walks are not parallel to each other.

\begin{figure}
\begin{center}
\includegraphics[angle=0,width=0.6\linewidth,clip=true]{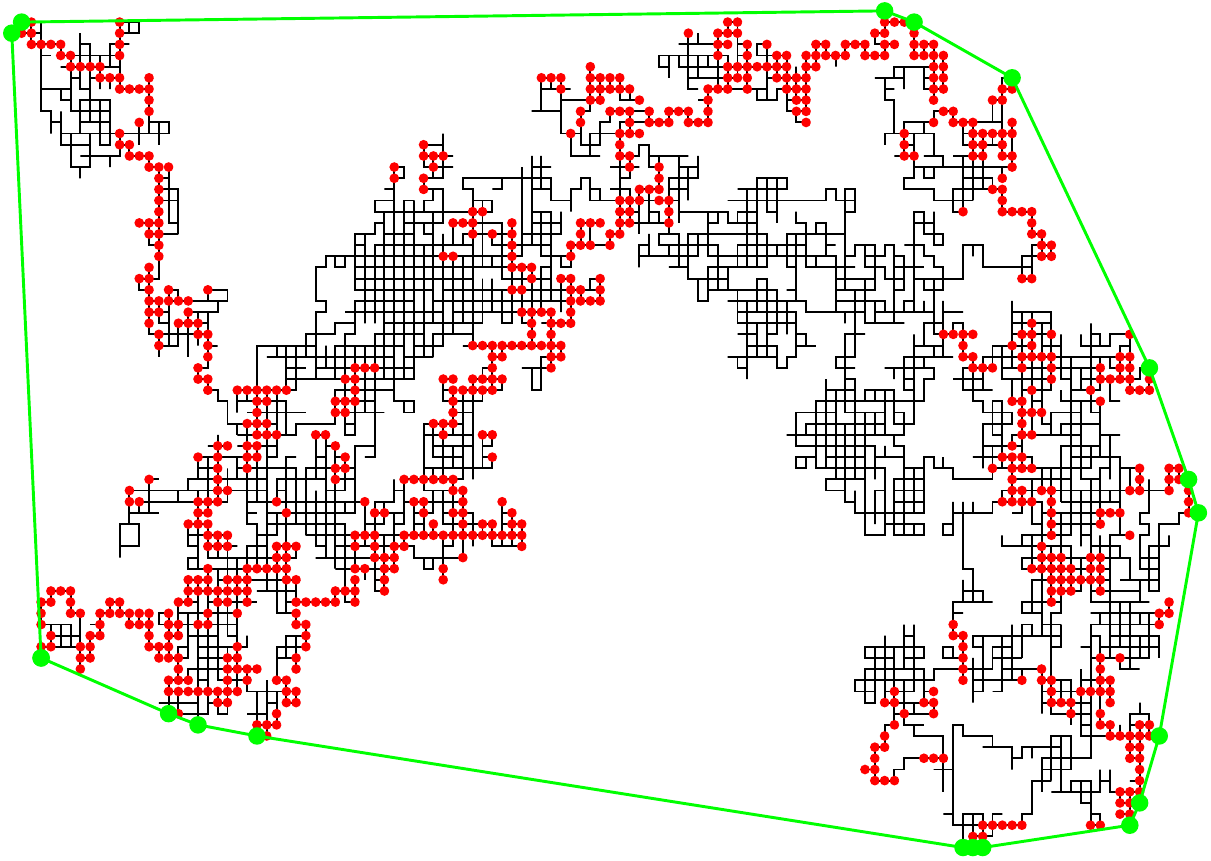}
\caption{\small
A P\'olya walk of 10,000 steps on the square lattice with $N=18$ and $R=820$.
Black line: trajectory of the random walker.
Green polygon: convex hull of the walk.
Green symbols: the 18 vertices of the convex hull.
Red symbols: the 802 other convex records.}
\label{plotpolya}
\end{center}
\end{figure}

The most notable difference with respect to the case of iid data points
investigated in section~\ref{iid} is that the position of the random walker spreads
away from the origin, according to the diffusion law~(\ref{xdiff}).
As a consequence, records are progressively buried deeper and deeper inside the current convex hull
which expands at the same diffusive scale as the walk.
In particular, the number $R_n$ of records is expected to grow much faster than $N_n$.
Stated otherwise, data points are by far not exchangeable,
so that the relation~(\ref{nriden}) can be expected to be violated by large amounts.
This is illustrated in figure~\ref{plotpolya},
showing a P\'olya walk of 10,000 steps on the square lattice such that $N=18$ and $R=820$.

Let us now turn to a quantitative analysis,
and consider first the number $N_n$ of vertices of the convex hull at time $n$.
For all microscopically isotropic planar random walks, including the Pearson walk,
the mean number $\mean{N_n}$ of vertices has been long known.
A combinatorial argument due to Baxter~\cite{B61}
yields the simple expression
\beq
\mean{N_n}=2H_n\approx2(\ln n+\gamma),
\label{bax}
\eeq
where the harmonic numbers $H_n$ are defined in~(\ref{hdef})
and $\gamma$ is Euler's constant.
No expression seems to be known for the corresponding variance.

\begin{figure}
\begin{center}
\includegraphics[angle=0,width=0.6\linewidth,clip=true]{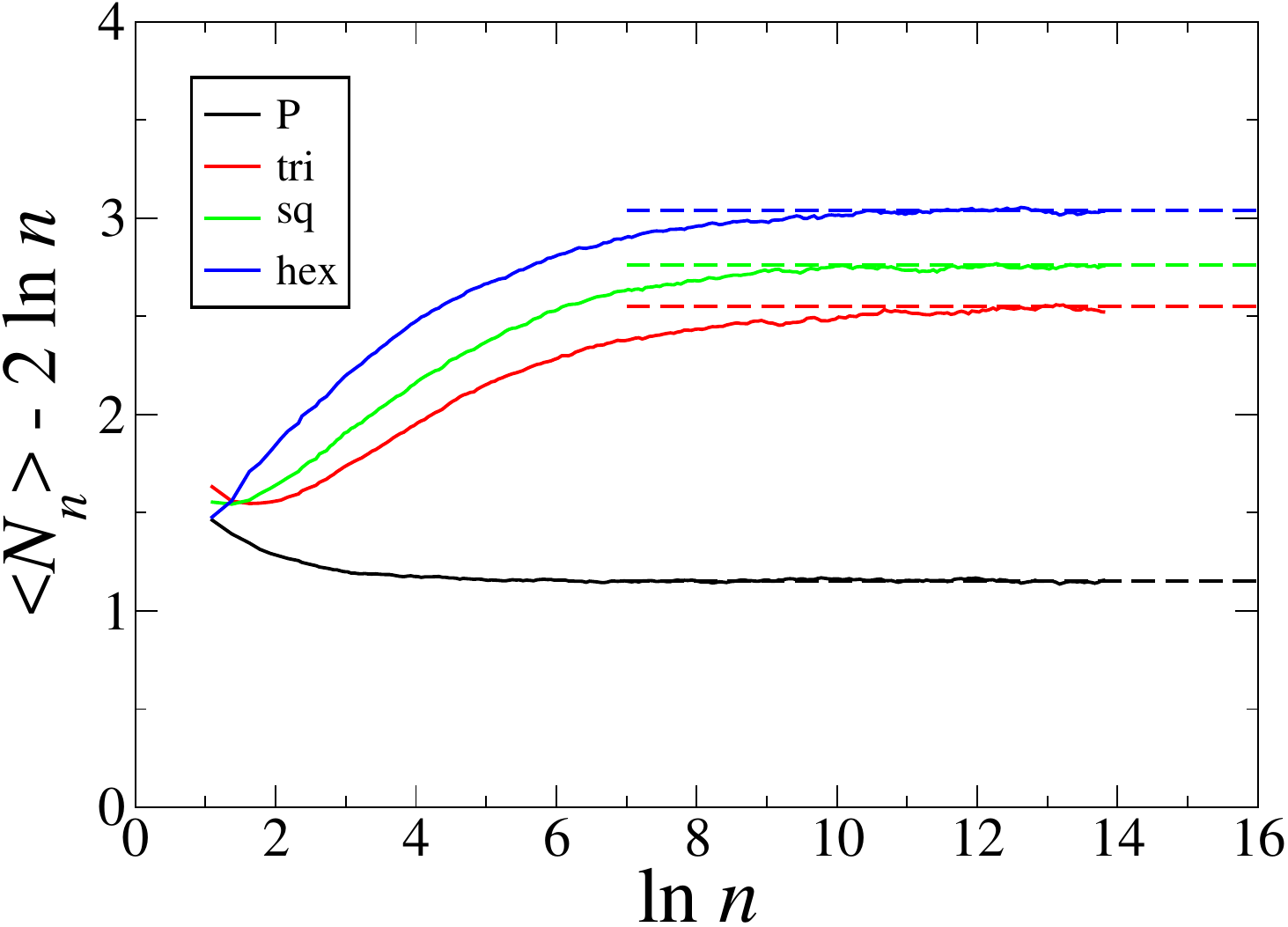}
\caption{\small
Difference $\mean{N_n}-2\ln n$ plotted against $\ln n$
for the four random walks considered in this work.
Black: Pearson walk.
Other colours: P\'olya walk on three lattices.
Red: triangular.
Green: square.
Blue: hexagonal.
Horizontal dashed lines: extrapolated limits yielding the values of
the finite part $G$ listed in table~\ref{csts}.}
\label{rwcst}
\end{center}
\end{figure}

The logarithmic growth law~(\ref{bax}) appears to be universal,
including its prefactor,
at least among the planar random walks we have investigated.
Figure~\ref{rwcst} shows plots of the difference $\mean{N_n}-2\ln n$ against $\ln n$
for the four examples of random walks considered in this work.
The plotted data strongly support the behaviour
\beq
\mean{N_n}\approx 2\ln n+G,
\label{rwnave}
\eeq
where the finite part $G$ depends on the type of walk (see table~\ref{csts} below).
For the Pearson walk,
data for finite $n$ converge rather fast from above to $G\approx1.15$,
in perfect agreement with the known limit
$G({\rm Pearson})=2\gamma=1.154431\dots$ (see~(\ref{bax})).
For the P\'olya walk on three lattices,
data exhibit a slower convergence from below to higher values of $G$.

\begin{figure}
\begin{center}
\includegraphics[angle=0,width=0.6\linewidth,clip=true]{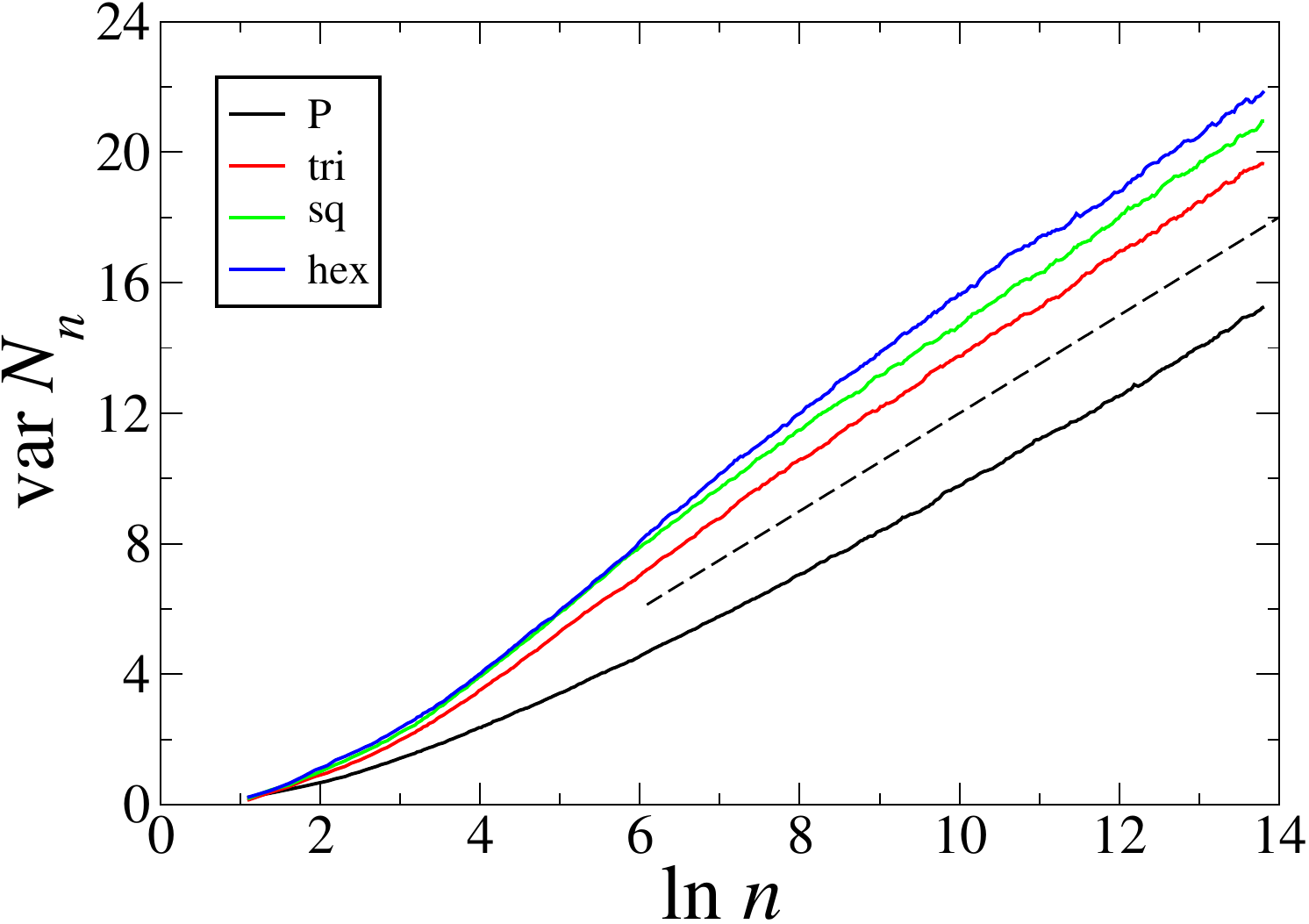}
\caption{\small
Variance of $N_n$ plotted against $\ln n$ for the four random walks (see legend).
Colours are as in figure~\ref{rwcst}.
The dashed line has slope 1.50.}
\label{rwvarn}
\end{center}
\end{figure}

Figure~\ref{rwvarn} shows plots of $\var N_n$ against $\ln n$ for the same four random walks.
The data support the universal logarithmic grow law
\beq
\var N_n\approx A_2\ln n,
\label{rwnvar}
\eeq
where the prefactor assumes the seemingly universal value
\beq
A_2\approx1.50,
\eeq
with an expected relative accuracy of a few percent.
The determination of the finite parts of the logarithmic law~(\ref{rwnvar})
is made difficult by the long transients in the data shown in figure~\ref{rwvarn}.

The results~(\ref{rwnave}) and~(\ref{rwnvar}) imply
that the distribution of $N_n$ is characterised by the universal limit Fano factor
\beq
F_N=\frac{A_2}{2}\approx0.75.
\eeq

Before we pursue,
it is worth emphasising the analogy between the convex records of planar random walks
and their radial records, investigated in~\cite{usplanar}.
The position $\x_n$ of the walker at time $n$
is a radial record if it is outside the circle whose radius\footnote{In~\cite{usplanar}
$n$ is denoted by $t$, $\bar r_n$ by $\bar R_t$, $R^\rad_n$ by $N_t$, and $B^\rad$ by $A$.}
\beq
\bar r_n=\max(\abs{\x_0},\dots,\abs{\x_{n-1}})
\label{rdef}
\eeq
is the largest distance to the origin reached by the walker before time $n$,
whereas~$\x_n$ is a convex record if it is outside the convex hull $\C(\x_0,\dots,\x_{n-1})$.
This convex polygon keeps forever fluctuating,
to the extent that it may assume any shape~\cite{RW18}.
It is nevertheless to be expected that the radius~$\bar r_n$
and the diameter of $\C(\x_0,\dots,\x_n)$
grow proportionally to the diffusive scale $\sqrt{n}$,
and so that there are similarities between the statistics of radial and convex records.

We recall that the main outcomes of~\cite{usplanar} concerning
radial records of random walks are based on the asymptotic equivalence
\beq
R^\rad_n\approx\frac{\bar r_n}{a}
\label{eqrad}
\eeq
between the number $R^\rad_n$ of radial records
and the radius $\bar r_n$ introduced in~(\ref{rdef}).
In~(\ref{eqrad}), the microscopic length scale $a$ depends on the type of walk,
whereas the radius $\bar r_n$ scales as
\beq
\bar r_n\approx U\sqrt{n},
\eeq
where the random variable $U$
has a universal distribution that is known explicitly~\cite[p.~280]{borodin}
and recalled in~\cite[Eq.~(4.12)]{usplanar}.

In particular, the mean number of radial records grows as
\beq
\mean{R^\rad_n}\approx B^\rad\sqrt{n},
\label{averrad}
\eeq
where the prefactor
\beq
B^\rad=\frac{\mean{U}}{a}
\eeq
depends on the type of walk through $a$.
Reference~\cite{usplanar} gives $B^\rad({\rm Pearson})\approx2.35$ for the Pearson walk
and $B^\rad({\rm square})\approx2.10$ for the P\'olya walk on the square lattice.
The number of radial records
is asymptotically distributed according to
\beq
\frac{R^\rad_n}{\mean{R^\rad_n}}\to X^\rad=\frac{U}{\mean{U}}.
\label{xrad}
\eeq
The limit random variable $X$ therefore has a universal distribution
such that $\mean{X^\rad}=1$, by construction, whereas
\beq
\var X^\rad=\frac{\var U}{\mean{U}^2}=0.110751\dots
\label{vxrad}
\eeq

\begin{figure}
\begin{center}
\includegraphics[angle=0,width=0.6\linewidth,clip=true]{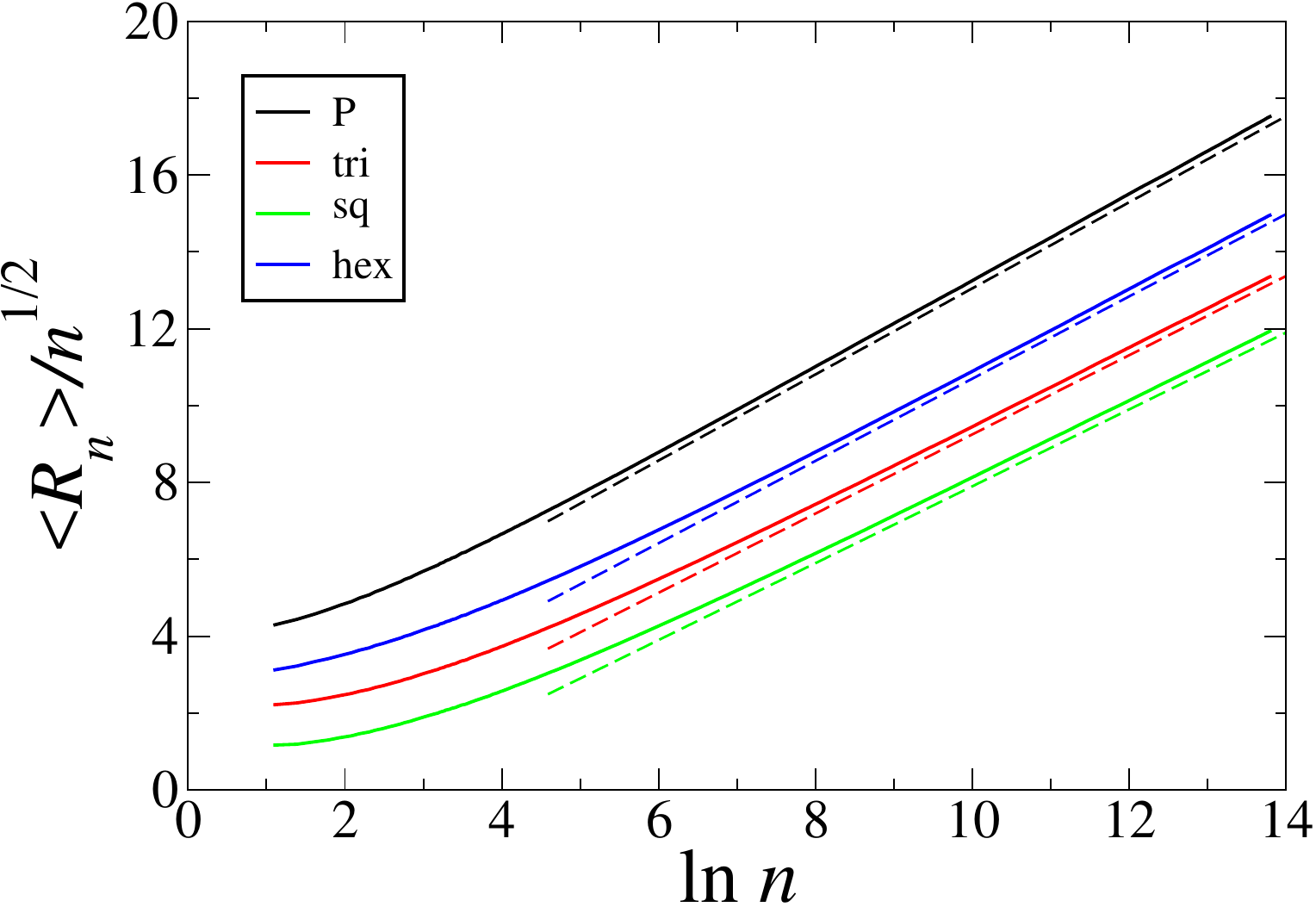}
\caption{\small
Ratio $\mean{R_n}/\sqrt{n}$ plotted against $\ln n$
for the four random walks (see legend).
Colours are as in figure~\ref{rwcst}.
Dashed lines have the slopes $B$ listed in table~\ref{csts}.}
\label{rwrave}
\end{center}
\end{figure}

Let us come back to convex records of random walks.
Concerning the mean number $\mean{R_n}$ of these records,
the numerical data shown in figure~\ref{rwrave} strongly suggest the growth law
\beq
\mean{R_n}\approx B\sqrt{n}\,\ln n,
\label{aver}
\eeq
where the prefactor $B$ has a weak dependence on the type of walk (see table~\ref{csts}).
A plausible justification for the above scaling law is that $R_n$ contains both
a factor~$\sqrt{n}$, already present in~(\ref{averrad}),
representing the diffusive growth of the walk,
and a factor~$\ln n$,
representing the number $N_n$ of vertices of the convex hull at the current time~$n$
(see~(\ref{rwnave})).

It is interesting to notice that the ratio
$B({\rm Pearson})/B({\rm square})\approx1.12$ equals
the corresponding ratio for radial records, i.e.,
$B^\rad({\rm Pearson})/B^\rad({\rm square})\approx2.35/2.10\approx1.12$,
given the available precision.
This coincidence suggests that the prefactor $B$, just as $B^\rad$,
only depends on the type of walk
through the microscopic length scale $a$ introduced in~\cite{usplanar}.

\begin{table}
\begin{center}
$$
\begin{array}{|l|c|c|c|}
\hline
\hbox{Type of walk} & z & G & B \\
\hline
\hbox{Pearson} & \infty & 1.15 & 1.12 \\
\hline
\hbox{P\'olya (triangular)} & 6 & 2.55 & 1.03 \\
\hbox{P\'olya (square)} & 4 & 2.76 & 1.00 \\
\hbox{P\'olya (hexagonal)} & 3 & 3.04 & 1.07 \\
\hline
\end{array}
$$
\caption{Various characteristic constants
of the four random walks considered in this work:
$z$ is the number of directions taken by the increments~$\del_n$,
i.e., the coordination number of the lattice,
$G$ is the finite part entering the logarithmic law~(\ref{rwnave})
for the mean number $\mean{N_n}$ of vertices,
$B$ is the prefactor of the growth law~(\ref{aver})
for the mean number $\mean{R_n}$ of convex records.}
\label{csts}
\end{center}
\end{table}

\begin{figure}
\begin{center}
\includegraphics[angle=0,width=0.6\linewidth,clip=true]{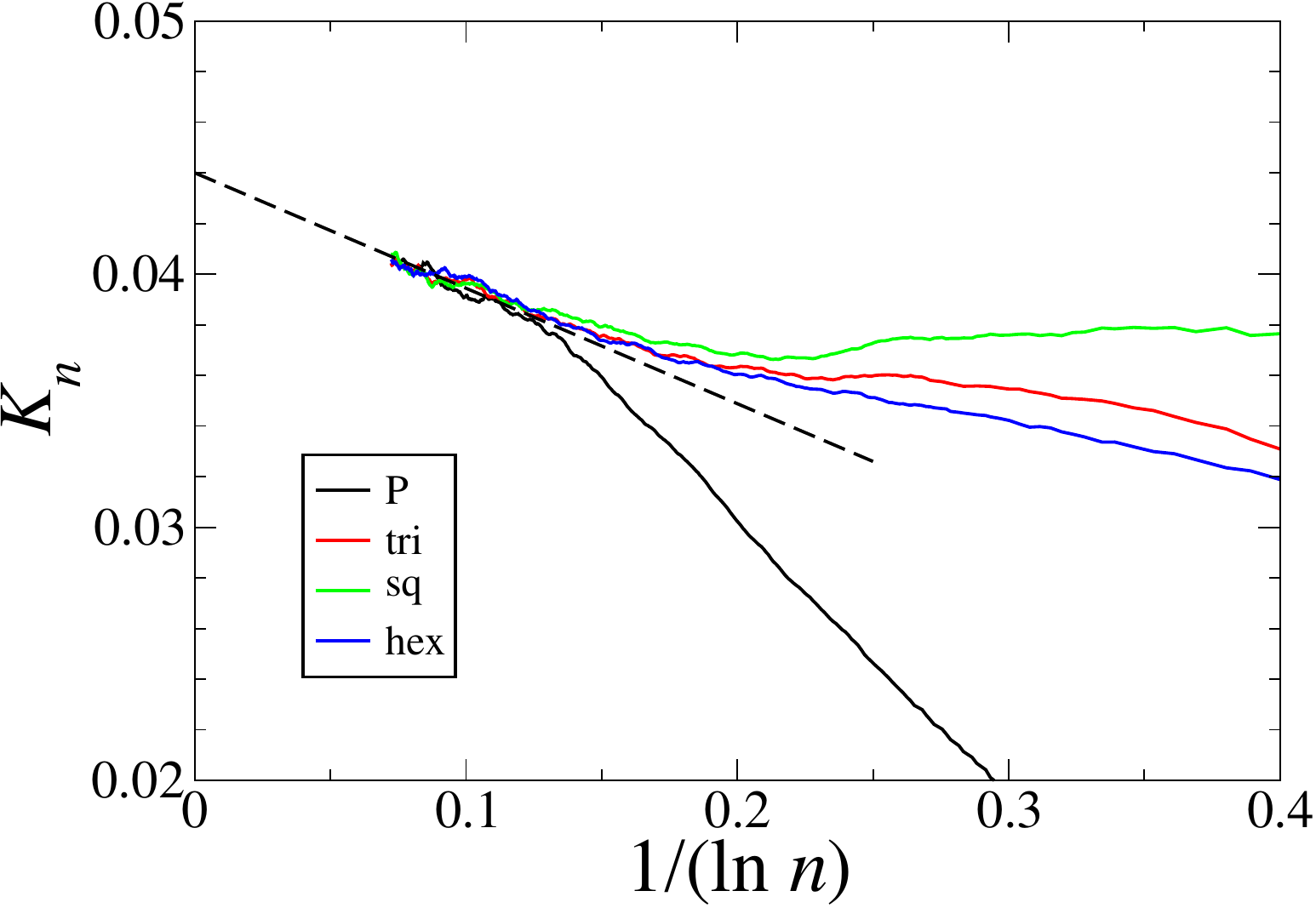}
\caption{\small
Reduced variance $K_n$ of the number of convex records plotted against $1/(\ln n)$,
for the four random walks (see legend).
Colours are as in figure~\ref{rwcst}.
The dashed line has intercept 0.044.}
\label{rwk}
\end{center}
\end{figure}

Concerning the full distribution of the number of convex records,
it is to be expected, in line with~(\ref{xrad}),
that $R_n$ keeps fluctuating
and is asymptotically distributed according to
\beq
\frac{R_n}{\mean{R_n}}\to X,
\eeq
where the limit random variable $X$ has a universal distribution such that $\mean{X}=1$.
This expectation, too, is corroborated by numerical simulations.
Figure~\ref{rwk} shows the reduced variance of the number of convex records,
\beq
K_n=\frac{\var R_n}{\mean{R_n}^2},
\eeq
plotted against $1/(\ln n)$ for the four random walks considered in this work.
The data is observed to converge linearly to the universal limit
\beq
K=\var X\approx0.044.
\label{vx}
\eeq
The very slowly decaying corrections in $1/(\ln n)$ however come as a surprise.

\begin{figure}
\begin{center}
\includegraphics[angle=0,width=0.6\linewidth,clip=true]{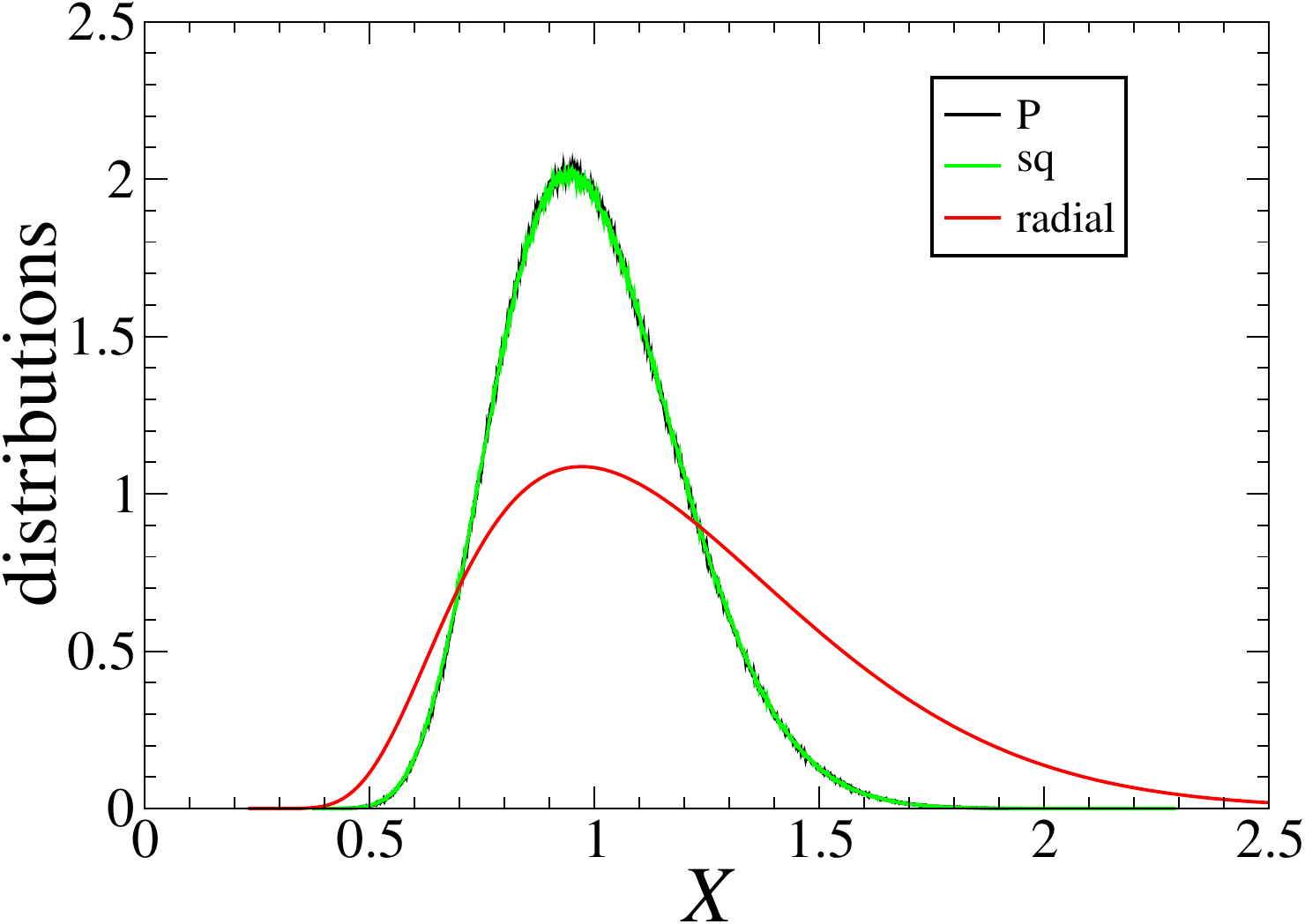}
\caption{\small
Distribution of the reduced variable $R_n/\mean{R_n}\approx X$
(unbinned data) for walks of $n=10,000$ steps.
Black: Pearson walk.
Green: P\'olya walk on the square lattice.
Red: exactly known distribution of $X^\rad$, corresponding to radial records (see~(\ref{xrad})).}
\label{histos}
\end{center}
\end{figure}

Figure~\ref{histos} shows the distribution of $R_n/\mean{R_n}$
for Pearson walk and for P\'olya walk on the square lattice with $n=10,000$ steps.
Both datasets fall on the same smooth curve,
representing a good approximation of the distribution of the limit variable $X$.
The exactly known distribution of $X^\rad$,
corresponding to radial records (see~(\ref{xrad})), is shown for comparison.
The latter distribution is broader.
The variance of $X^\rad$ is indeed some 2.5 times larger than that of $X$
(see~(\ref{vxrad}),~(\ref{vx})).

\newpage

\section{Discussion}
\label{disc}

The primary aim of this paper is to draw the reader's attention to convex records.
Among all possible definitions of multivariate records,
the convex records investigated in this work
stand out for the elegance of their geometric definition
and for the ensuing invariance of their construction under the affine group.

The present work is focused on the bivariate (i.e., two-dimensional) case.
This choice is motivated by simplicity,
and chiefly by the existence of a simple algorithm to recursively build convex hulls
of growing data sets.
We wish to highlight that some of the statistics of higher-dimensional convex records
can be sketched in the light of the results presented above.
To be more specific,
for iid (independent and identically distributed) data points,
the identity~(\ref{nriden}) has been instrumental in relating
the mean number $\mean{R_n}$ of convex records up to time~$n$
to the mean number $\mean{N_n}$ of vertices of the convex hull of the first $n$ points.
The above identity is in fact quite general,
and applies to iid data points in any dimension $d$.
Let us consider two characteristic examples,
for which some results on convex hulls are available in the mathematical literature.
For uniform data points in an arbitrary $d$-dimensional convex polytope, we have
\beq
\mean{N_n}\approx A_1(\ln n)^{d-1},
\eeq
where the prefactor $A_1$ is known~\cite{BB93},
and so~(\ref{nriden}) yields
\beq
\mean{R_n}\approx B_1(\ln n)^d,\qquad B_1=\frac{A_1}{d}.
\eeq
For uniform data points in a $d$-dimensional sphere, we have
\beq
\mean{N_n}\approx A_1\,n^{(d-1)/(d+1)},
\eeq
where $A_1$ is also known exactly~\cite{R70,R03},
and so~(\ref{nriden}) yields
\beq
\mean{R_n}\approx B_1\,n^{(d-1)/(d+1)},\qquad B_1=\frac{d+1}{d-1}A_1.
\eeq
Moreover, it is quite plausible that the variances of $N_n$ and $R_n$ generically grow
proportionally to the corresponding mean values given above,
resulting in finite limit Fano factors $F_N$ and $F_R$,
such as those shown in figure~\ref{plotfanos}.
On the other hand, for isotropic random walks in $d$ dimensions,
a known exact expression of $\mean{N_n}$ for any finite number $n$ of steps~\cite{K17}
reads asymptotically
\beq
\mean{N_n}\approx\frac{2}{(d-1)!}\,(\ln n)^{d-1}.
\eeq
We can therefore expect that the mean number of convex records scales as
\beq
\mean{R_n}\approx B\sqrt{n}\,(\ln n)^{d-1}.
\eeq
This estimate should hold for the Pearson walk
and for the P\'olya walk on a lattice in any dimension $d$,
with the prefactor $B$ depending on the type of walk.
Finally, it is also to be expected that $R_n/\mean{R_n}$ keeps fluctuating
and goes to a universal random variable $X$, whose distribution only depends on $d$.

We conclude with a discussion on the growth of the number of records
for various kinds of records in sequences of $d$-dimensional iid data points.
The mean number of records
can be expressed in terms of the record-breaking probability~$Q_n$ (see~(\ref{rpro})).
For the convex records studied in this work,
a $d$-dimensional simplex has $d+1$ vertices.
As a result, we have generically $n Q_n=\mean{N_n}\ge d+1$ for all $n\ge d+1$,
implying that the mean number of records, $\mean{R_n}$, grows at least as $(d+1)\ln n$.
This minimal logarithmic growth should be compared with the growth rates observed for other definitions of multivariate records.
Consider simultaneous records (also known as complete or concomitant records),
where there is a record at time $n$ if each component~$x_n^i$ of $\x_n$
is larger than all previous~$x_m^i$.
There, the record-breaking probability $Q_n$ can assume any value between 0 and $1/n$
(see~\cite{gnedin} and references therein).
The upper bound coincides with the result~(\ref{pusuel}) of the univariate case.
Simple explicit examples can be built in any dimension $d$,
for which either $Q_n=0$ for all $n>1$, or $Q_n=1/n$.
For data points with independent components $x_n^i$ following continuous distributions,
we have $Q_n=1/n^d$.
For Gaussian data points, we have $Q_n\sim n^{-\alpha}$,
where the exponent $\alpha>1$ depends continuously on parameters.
In both examples,
the total number of simultaneous records remains finite for an infinitely large dataset.

\ack

It is a pleasure to thank Philippe Naveau for stimulating discussions
which motivated the present work.

\section*{Data availability statement}

Data sharing not applicable to this article.

\section*{Conflict of interest}

The authors declare no conflict of interest.

\section*{Orcid ids}

Claude Godr\`eche https://orcid.org/0000-0002-1833-3490

\noindent Jean-Marc Luck https://orcid.org/0000-0003-2151-5057

\appendix

\section{The distribution of univariate records}
\label{app}

This appendix is a self-contained reminder of the classical theory
of the statistics of records in sequences of iid univariate random variables
drawn from an arbitrary continuous distribution
(see~\cite{chandler,renyi1,renyi2,glick,arn,nevzorov,nevzorov2,bunge}).
In this setting,
there is a record at time $n$ with probability (see~(\ref{pusuel}))
\beq
Q_n=\frac{1}{n},
\eeq
and the occurrences of records at different times are statistically independent.

The quantity of interest is the number $R_n$ of records up to time $n$.
The distribution of this random number,
\beq
p_n(k)=\prob(R_n=k)\qquad(k=1,\dots,n),
\eeq
is conveniently encoded in the generating function
\beq
G_n(z)=\mean{z^{R_n}}=\sum_{k=1}^n p_n(k)z^k,
\label{gdef}
\eeq
which is a polynomial in $z$ with degree $n$.
The independence of the occurrences of records at different times yields the product formula
\beq
G_n(z)=\prod_{m=1}^n(1-Q_m+z Q_m)=\prod_{m=1}^n\frac{m-1+z}{m}.
\eeq
This can be recast as
\beq
G_n(z)=\frac{\Gamma(n+z)}{n!\,\Gamma(z)}=\frac{1}{n!}\sum_{k=1}^n\stir{n}{k}z^k,
\label{1dg}
\eeq
where the $\stir{n}{k}$ are the Stirling numbers of the first kind,
which are ubiquitous in combinatorics (see e.g.~\cite{GKP,FS}).
The distribution of $R_n$ therefore reads
\beq
p_n(k)=\frac{1}{n!}\stir{n}{k}.
\eeq
The integer $\stir{n}{k}$ is, among many other things,
the number of permutations of $n$ objects having $k$ cycles,
so that the number $R_n$ of records is distributed as the number of cycles
in a uniform random permutation.

In particular, the mean value and the variance of $R_n$ read
\beqa
\mean{R_n}=\sum_{m=1}^n Q_m=H_n\approx\ln n+\gamma,
\label{1dave}
\\
\var R_n=\sum_{m=1}^n Q_m(1-Q_m)=H_n-H_n^{(2)}\approx\ln n+\gamma-\frac{\pi^2}{6},
\label{1dvar}
\eeqa
where
\beq
H_n=\sum_{m=1}^n\frac{1}{m},\qquad
H_n^{(2)}=\sum_{m=1}^n\frac{1}{m^2},
\label{hdef}
\eeq
and $\gamma$ is Euler's constant.

The number $R_n$ of records takes its smallest and largest values with respective probabilities
\beq
p_n(1)=\frac{1}{n},\qquad
p_n(n)=\frac{1}{n!}.
\label{unipex}
\eeq

The full distribution of $R_n$ takes a simple asymptotic form at large $n$.
A first approximation to~(\ref{1dg}) at large $n$ reads
\beq
G_n(z)\sim\e^{(z-1)\ln n},
\eeq
and so the distribution of $R_n$ becomes a Poisson distribution
with parameter $\lambda=\ln n$ and Fano factor $F=1$.
A more refined asymptotic form of~(\ref{1dg}) is
\beq
G_n(z)\approx\frac{\e^{(z-1)\ln n}}{\Gamma(z)},
\eeq
implying that the cumulants of $R_n$ grow as
\beq
\mean{R_n^p}_c\approx\ln n+a_p,
\label{1dcum}
\eeq
with a common logarithmic term with unit prefactor,
and finite parts $a_p$ are given by
\beq
\sum_{p\ge1}\frac{a_p}{p!}\,s^p=-\ln\Gamma(\e^s),
\eeq
i.e.,
\beqa
a_1=\gamma,\qquad
a_2=\gamma-\frac{\pi^2}{6},\qquad
a_3=\gamma-\frac{\pi^2}{2}+2\zeta(3),
\nonumber\\
a_4=\gamma-\frac{7\pi^2}{6}-\frac{\pi^4}{15}+12\zeta(3),
\eeqa
and so on.
The first two expressions agree with~(\ref{1dave}) and~(\ref{1dvar}).

\section*{References}

\bibliography{paper.bib}

\begin{thebibliography}{10}

\bibitem{revue}
C.~Godr\`eche, S.~N. Majumdar, and G.~Schehr.
\newblock Record statistics of a strongly correlated time series: random walks
  and \protect{L\'evy} flights.
\newblock {\em J. Phys. A: Math. Theor.}, 50:333001, 2017.

\bibitem{chandler}
K.~N. Chandler.
\newblock The distribution and frequency of record values.
\newblock {\em J. R. Statist. Soc. B.}, 14:220--228, 1952.

\bibitem{renyi1}
A.~R\'enyi.
\newblock Th\'eorie des \'el\'ements saillants d'une suite d'observations.
\newblock {\em Ann. Sci. Univ. Clermont-Ferrand}, 8:7--13, 1962.

\bibitem{renyi2}
A.~R\'enyi.
\newblock Th\'eorie des \'el\'ements saillants d'une suite d'observations.
\newblock In {\em Colloquium on Combina\-torial Methods in Probability Theory},
  pages 104--117. Mathematical Institute of Aarhus University, Aarhus, Denmark,
  1962.

\bibitem{glick}
N.~Glick.
\newblock Breaking records and breaking boards.
\newblock {\em Amer. Math. Monthly}, 85:2--26, 1978.

\bibitem{arn}
B.~C. Arnold, N.~Balakrishnan, and H.~N. Nagaraja.
\newblock {\em Records}.
\newblock Wiley, New York, 1998.

\bibitem{nevzorov}
V.~B. Nevzorov and N.~Balakrishnan.
\newblock A record of records.
\newblock {\em Handbook of Statistics}, 16:515--570, 1998.

\bibitem{nevzorov2}
V.~B. Nevzorov.
\newblock {\em Records: Mathematical Theory (Translation of Mathematical
  Monographs vol {\bf 194})}.
\newblock American Mathematical Society, Providence, RI, 2001.

\bibitem{bunge}
J.~Bunge and C.~M. Goldie.
\newblock Record sequences and their applications.
\newblock {\em Handbook of Statistics}, 19:277--308, 2001.

\bibitem{feller2}
W.~Feller.
\newblock {\em An Introduction to Probability Theory and its Applications},
  volume~2.
\newblock Wiley, New York, 2nd edition, 1971.

\bibitem{K66}
M.~G. Kendall.
\newblock Discrimination and classification, multivariate analysis.
\newblock In P.~R. Krishnaiah, editor, {\em Multivariate Analysis}, pages
  165--184, New York, 1966. Academic.

\bibitem{B76}
V.~Barnett.
\newblock The ordering of multivariate data.
\newblock {\em J. R. Statist. Soc. A}, 139:318--355, 1976.

\bibitem{GR89}
C.~M. Goldie and S.~I. Resnick.
\newblock Records in a partially ordered set.
\newblock {\em Ann. Prob.}, 17:678--699, 1989.

\bibitem{GR95}
C.~M. Goldie and S.~I. Resnick.
\newblock Many multivariate records.
\newblock {\em Stoch. Proc. Appl.}, 59:185--216, 1995.

\bibitem{gnedin}
A.~V. Gnedin.
\newblock Records from a multivariate normal sample.
\newblock {\em Stat. Prob. Lett.}, 39:11--15, 1998.

\bibitem{HT10}
H.~K. Hwang and T.~H. Tsai.
\newblock Multivariate records based on dominance.
\newblock {\em Electron. J. Prob.}, 15:1863--1892, 2010.

\bibitem{DZ18}
C.~Dombry and M.~Zott.
\newblock Multivariate records and hitting scenarios.
\newblock {\em Extremes}, 21:343--361, 2018.

\bibitem{BSN20}
N.~Balakrishnan, A.~Stepanov, and V.~B. Nevzorov.
\newblock North-east bivariate records.
\newblock {\em Metrika}, 83:961--976, 2020.

\bibitem{TF21}
S.~Tat and M.~R. Faridrohani.
\newblock A new type of multivariate records: depth-based records.
\newblock {\em Statistics}, 55:296--320, 2021.

\bibitem{K95}
M.~Kaluszka.
\newblock Estimates of some probabilities in multidimensional convex records.
\newblock {\em Applicationes Math.}, 23:1--11, 1995.

\bibitem{KM}
M.~G. Kendall and P.~A.~P. Moran.
\newblock Geometrical probability.
\newblock In M.~G. Kendall, editor, {\em Griffin's Statistical Monographs and
  Courses}, volume~10. Hafner, New York, 1963.

\bibitem{SW}
R.~Schneider and W.~Weil.
\newblock {\em Stochastic and Integral Geometry}.
\newblock Springer, Berlin, 2008.

\bibitem{MCR}
S.~N. Majumdar, A.~Comtet, and J.~Randon-Furling.
\newblock Random convex hulls and extreme value statistics.
\newblock {\em J. Stat. Phys.}, 138:955--1009, 2010.

\bibitem{sylvester}
J.~J. Sylvester.
\newblock Problem 1491.
\newblock {\em The Educational Times}, April 1864.

\bibitem{P89}
R.~E. Pfeifer.
\newblock The historical development of \protect{J. J. Sylvester's} four point
  problem.
\newblock {\em Math. Mag.}, 62:309--317, 1989.

\bibitem{RS63}
A.~R\'enyi and R.~Sulanke.
\newblock {\"U}ber die konvexe \protect{H\"ulle} von $n$ zuf\"allig gew\"ahlten
  \protect{Punkten}.
\newblock {\em Z. Wahr.}, 2:75--84, 1963.

\bibitem{E65}
B.~Efron.
\newblock The convex hull of a random set of points.
\newblock {\em Biometrika}, 52:331--343, 1965.

\bibitem{G88}
P.~Groeneboom.
\newblock Limit theorems for convex hulls.
\newblock {\em Probab. Th. Rel. Fields}, 79:327--368, 1988.

\bibitem{fano}
U.~Fano.
\newblock Ionization yield of radiations. \protect{II. The} fluctuations of the
  number of ions.
\newblock {\em Phys. Rev.}, 72:26--29, 1947.

\bibitem{TTT}
J.~Tworzyd\l{o}, B.~Trauzettel, M.~Titov, A.~Rycerz, and C.~W.~J. Beenakker.
\newblock \protect{Sub-Poissonian} shot noise in graphene.
\newblock {\em Phys. Rev. Lett.}, 96:246802, 2006.

\bibitem{mandel}
L.~Mandel.
\newblock \protect{Sub-Poissonian} photon statistics in resonance fluorescence.
\newblock {\em Optics Lett.}, 4:205--207, 1979.

\bibitem{B06}
C.~Buchta.
\newblock The exact distribution of the number of vertices of a random convex
  chain.
\newblock {\em Mathematika}, 53:247--254, 2006.

\bibitem{B09}
C.~Buchta.
\newblock On the number of vertices of the convex hull of random points in a
  square and a triangle.
\newblock {\em Anzeiger Abt. II}, 143:3--10, 2009.

\bibitem{M17}
J.~F. Marckert.
\newblock The probability that $n$ random points in a disk are in convex
  position.
\newblock {\em Braz. J. Prob. Stat.}, 31:320--337, 2017.

\bibitem{FH04}
S.~Finch and I.~Hueter.
\newblock Random convex hulls: a variance revisited.
\newblock {\em Adv. Appl. Prob.}, 36:981--986, 2004.

\bibitem{C70}
H.~Carnal.
\newblock Die konvexe \protect{H\"ulle} von $n$ rotationssymmetrisch verteilten
  \protect{Punkten}.
\newblock {\em Z. Wahr.}, 15:168--176, 1970.

\bibitem{H94}
I.~Hueter.
\newblock The convex hull of a normal sample.
\newblock {\em Adv. Appl. Prob.}, 26:855--875, 1994.

\bibitem{H99}
I.~Hueter.
\newblock Limit theorems for the convex hull of random points in higher
  dimensions.
\newblock {\em Trans. Amer. Math. Soc.}, 351:4337--4363, 1999.

\bibitem{KL}
P.~L. Krapivsky and J.~M. Luck.
\newblock On multidimensional record patterns.
\newblock {\em J. Stat. Mech.}, 2020:063205, 2020.

\bibitem{A91}
D.~J. Aldous, B.~Fristedt, P.~S. Griffin, and W.~E. Pruitt.
\newblock The number of extreme points in the convex hull of a random sample.
\newblock {\em J. Appl. Prob.}, 28:287--304, 1991.

\bibitem{M00}
B.~Mass\'e.
\newblock On the \protect{LLN} fo the number of vertices of a random convex
  hull.
\newblock {\em Adv. Appl. Prob.}, 32:675--681, 2000.

\bibitem{reed}
W.~J. Reed.
\newblock Random points in a simplex.
\newblock {\em Pacific J. Math.}, 54:183--198, 1974.

\bibitem{alagar}
V.~S. Alagar.
\newblock On the distribution of a random triangle.
\newblock {\em J. Appl. Prob.}, 14:284--297, 1977.

\bibitem{V95}
P.~Valtr.
\newblock Probability that $n$ random points are in convex position.
\newblock {\em Discrete Comp. Geom.}, 13:637--643, 1995.

\bibitem{V96}
P.~Valtr.
\newblock The probability that $n$ random points in a triangle are in convex
  position.
\newblock {\em Combinatorica}, 16:567--573, 1996.

\bibitem{HCS08}
H.~J. Hilhorst, P.~Calka, and G.~Schehr.
\newblock \protect{Sylvester's} question and the random acceleration process.
\newblock {\em J. Stat. Mech.}, 2008:P10010, 2008.

\bibitem{Pearson}
K.~Pearson.
\newblock The problem of the random walk.
\newblock {\em Nature}, 72:294, 1905.

\bibitem{Polya}
G.~P\'olya.
\newblock Wahrscheinlichkeitstheoretisches \"uber die \protect{Irrfahrt}.
\newblock {\em Mitt. der Phys. Ges. Z\"urich}, 19:75--86, 1919.

\bibitem{SW61}
F.~Spitzer and H.~Widom.
\newblock The circumference of a convex polygon.
\newblock {\em Proc. Amer. Math. Soc.}, 12:506--509, 1961.

\bibitem{B61}
G.~Baxter.
\newblock A combinatorial lemma for complex numbers.
\newblock {\em Ann. Math. Statist.}, 32:901--904, 1961.

\bibitem{T80}
L.~Takacs.
\newblock Expected perimeter length.
\newblock {\em Amer. Math. Monthly}, 87:142, 1980.

\bibitem{S02}
J.~M. Steele.
\newblock The \protect{Bohnenblust-Spitzer} algorithm and its applications.
\newblock {\em J. Comp. Appl. Math.}, 142:235--249, 2002.

\bibitem{K17}
Z.~Kabluchko, V.~Vysotsky, and D.~Zaporozhets.
\newblock Convex hulls of random walks: Expected number of faces and face
  probabilities.
\newblock {\em Adv. Math.}, 320:595--629, 2017.

\bibitem{usplanar}
C.~Godr\`eche and J.~M. Luck.
\newblock On sequences of records generated by planar random walks.
\newblock {\em J. Phys. A: Math. Theor.}, 54:325003, 2021.

\bibitem{RW18}
J.~McRedmond and A.~R. Wade.
\newblock The convex hull of a planar random walk: perimeter, diameter, and
  shape.
\newblock {\em Electron. J. Prob.}, 23:1--24, 2018.

\bibitem{borodin}
A.~N. Borodin and P.~Salminen.
\newblock {\em Handbook of Brownian Motion - Facts and Formulae}.
\newblock Birkh\"auser, Basel, 1996.

\bibitem{BB93}
I.~B\'ar\'any and C.~Buchta.
\newblock Random polytopes in a convex polytope, independence of shape, and
  concentration of vertices.
\newblock {\em Math. Ann.}, 297:467--497, 1993.

\bibitem{R70}
H.~Raynaud.
\newblock Sur l'enveloppe convexe des nuages de points al\'eatoires dans
  \protect{${\mathbb{R}}^n$}.
\newblock {\em J. Appl. Prob.}, 7:35--48, 1970.

\bibitem{R03}
M.~Reitzner.
\newblock Random polytopes and the \protect{Efron-Stein} jackknife inequality.
\newblock {\em Ann. Prob.}, 31:2136--2166, 2003.

\bibitem{GKP}
R.~L. Graham, D.~E. Knuth, and O.~Patashnik.
\newblock {\em Concrete Mathematics: A Foundation for Computer Science}.
\newblock Addison-Wesley, Reading, MA, 1989.

\bibitem{FS}
P.~Flajolet and R.~Sedgewick.
\newblock {\em Analytic Combinatorics}.
\newblock Cambridge University Press, Cambridge, 2009.

\end{thebibliography}

\end{document}